# Guideline for Manual Process Discovery in Industrial IoT

Linda Kölbel, Markus Hornsteiner, Stefan Schönig
University of Regensburg, Universitätsstraße 31, 93053 Regensburg, Bavaria, Germany

**In industry, the networking and automation of machines through the Internet of Things (IoT) continues to increase, leading to greater digitalization of production processes. Traditionally, business and production processes are controlled, optimized and monitored using business process management methods that require process discovery. However, these methods cannot be fully applied to industrial production processes. Nevertheless, processes in the industry must also be monitored and discovered for this purpose.**
**The aim of this paper is to develop an approach for process discovery methods and to adapt existing process discovery methods for application to industrial processes. The adaptations of classic discovery methods are presented as universally applicable guidelines specifically for the Industrial Internet of Things (IIoT). In order to create an optimal process model based on process evaluation, different methods are combined into a standardized discovery approach that is both efficient and cost-effective.**

***Keywords:*** *Business Process Management; Process Discovery; Industrial Internet of Things; Guideline Development*

## 1 Introduction

The Industrial Internet of Things (IIoT) connects people with machines and machines with each other, promoting faster and more efficient processes in the business context (Boyes et al. 2018). To optimize these processes, companies practice Business Process Management (BPM), which employs various methods to identify, control, analyze, optimize, and monitor business processes (Dumas et al. 2021).
For companies to continue benefiting from BPM in the era of IIoT, they need methods that apply BPM principles to industrial processes. The first step in BPM is process discovery, which involves identifying and documenting processes (Dumas et al. 2021). However, since BPM has traditionally focused on business processes, existing methods for process discovery need to be adjusted and developed to suit industrial contexts.
Due to the digital evolution of business processes within the framework of IIoT, some classical BPM techniques can no longer be applied in their original form. To address this problem, BPM techniques must be revised and adapted to meet the requirements of IIoT, capturing process flows, data generated and used during processes, and communication between machines.
Classical process management discovery methods work well for manual processes with few automated components. Fully automated processes can be discovered using process mining techniques, which analyze machine-generated log data. However, partially automated processes that combine manual and automated activities pose a challenge, as process mining techniques are ineffective due to the absence of log data for manual activities.To discover such processes in the IIoT, this paper proposes adapting classical discovery methods and documenting these adaptations as guidelines. The discovery of a process should be conducted from different perspectives to ensure a comprehensive view. The resulting process model should be complete and correct, encompassing the process flow, data flow, and machine communication, addressing all aspects relevant to the IIoT.

The structure of this paper is as follows: In Section 2, we present the research method and its application. In Section 3, we explain the fundamental background of the contexts used in this paper. In Section 4, we discuss the related work, how it was discovered, and its influence on this paper. In Section 5, we introduce the developed guidelines and their application. In Section 6 we present a combining example of the different methods. In Section 7, these guidelines are applied to specific use cases. In Section 8, we evaluate the guidelines and their effectiveness. Finally, in Section 9, we conclude with the limitations and possible directions for future research.

## 2 Research Method

In this paper, the Design Science Research (DSR) approach of Hevner et al. (2004) is utilized, as it aims to develop a suitable solution, called artifact, for a specific problem in the field of information systems. The approach consists of seven guidelines, which are explained below and their implementation in the context of this paper is discussed. However, the order and type of implementation is not strictly defined in order not to restrict the creativity of the researchers.

**Design as an Artifact** states that the research process must lead to an artifact, such as a model or a method. The artifact of this work is a novel method for manual discovery of IIoT processes that is described in detail and applicable in practice.

**Problem Relevance** requires that the problem is not only relevant academically, but also in industry. Since methods for discovering IIoT processes do not yet exist, but the importance of IIoT in industry is constantly increasing and the discovery of such processes using traditional methods is becoming increasingly complex. The developed artifact should extend and adapt established BPM methods to systematically discover IIoT-based processes.

**Design Evaluation** requires that the artifact to be developed is comprehensively evaluated. This evaluation of the artifact developed in this paper for its understandability and usability is discussed in Section 8. The evaluation is divided into two parts and consists of an application of the artifacts in the form of a use case and a survey using questionnaires on understandability and usability.

**Research Contribution** requires that the artifact to be developed provides a novel contribution to the state of the art. In order to determine the state of the art, a systematic literature search was carried out, which is discussed in Section 4. No published scientific papers addressing manual discovery in IIoT processes could be found. Therefore, the artifact developed in this paper represents a novel research contribution and fulfills this guideline.

**Research Rigor** requires that the artifact is developed using accepted and therefore reproducible scientific methods. For this purpose, this paper utilizes the already described DSR methodology according to Hevner et al. (2004) for the development of the artifact. In addition, a literature review followed the guideline of Okoli and Schabram (2010) is carried out in Section 4, on the results of which the artifact developed in this paper is based.

**Design as a Search Process** makes it clear that the development of the artifact is not a one-off event, but a process with several iterations. This was also the case in the development of the artifact presented in this paper. Based on the literature, several iterations of the artifact were created and continuously improved to the current status using, for example, procedures from various research disciplines and expert feedback.

**Communication of Research** requires that the artifact be made accessible to an informed and interested public. Usually in the form of a publication in a specialist medium. Accordingly, the publication of the artifact of this paper in a journal. Figure 1 shows an outline of the phases and information described in this section.

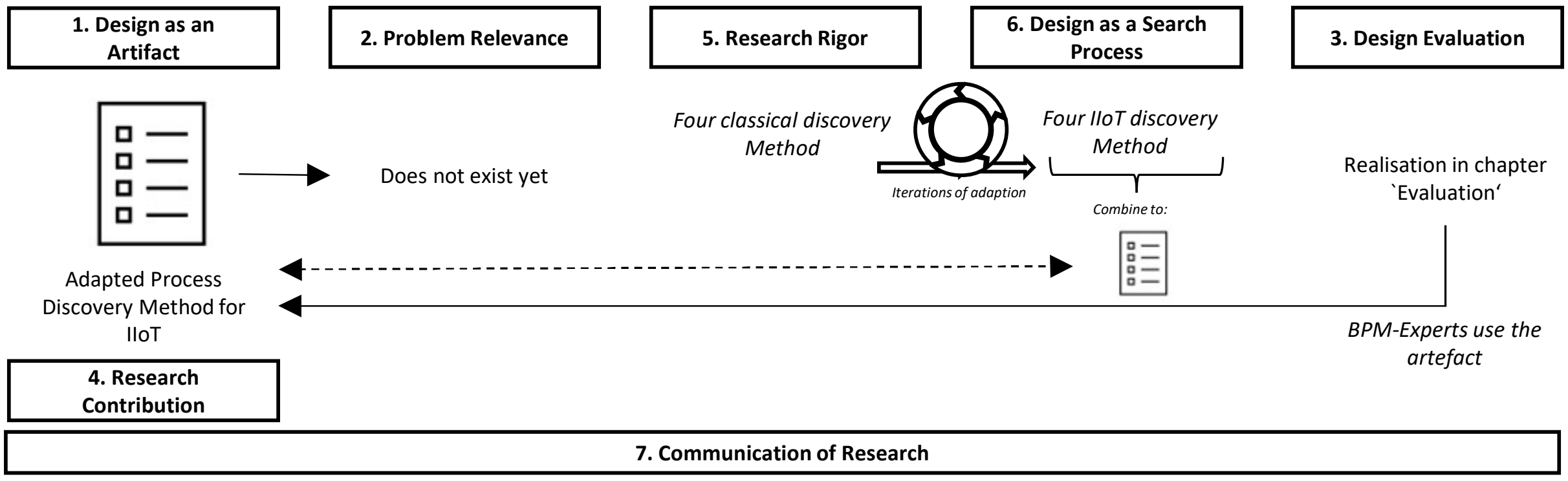

Figure 1: Application of the Design-Phases of DSR in the present work.

# 3 Background

## 3.1 Business Process Management and Process Discovery

BPM encompasses all tasks and measures to make processes more efficient and effective (Hansen et al. 2019). BPM should serve as a decision-making aid for process improvement and support the management of companies (Weske 2012). In particular, the aim is to shorten throughput times, increase efficiency, save costs and minimize error rates, which then contributes to increasing competitiveness (Dumas et al. 2021; Bernardo et al. 2017). BPM is also seen as a strategy for gaining a competitive advantage, whereby numerous definitions exist (zur Muehlen and Ho 2005). This paper is based on the definition by Dumas et al. (2021):

*A body of methods, techniques, and tools to identify, discover, analyze, redesign, execute, and monitor business processes in order to optimize their performance.*

This paper focuses on process discovery and explicitly addresses the concepts, methods, and techniques involved. It involves gathering information about the current state of a process, typically through methods such as document analysis, observation, interviews, or workshops (Dumas et al. 2021; Hansen et al. 2019; Gronau 2017). Documents containing relevant information include organizational structures, workflows, responsibilities and technical aspects (Dumas et al. 2021; Becker et al. 2012). Challenges include division of labor, terminology inconsistencies, and process complexity, often requiring multiple iterations to resolve ambiguities (Hansen et al. 2019). Modeling in Event-driven process chains (EPC), Unified Modeling Language (UML), or Business Process Modeling Notation (BPMN) , for example, helps with process analysis and optimization and serves as a knowledge base for training (Jakobs and Spanke 2011).

## 3.2 Industrial IoT

The IIoT is an industry-oriented form of the IoT. The IoT connects diverse, traditionally non-networked devices and collects data for analysis. Each object is accessible through a unique address and can communicate with other objects using standardized protocols. The difference between the IIoT and the IoT is primarily in the technologies used and the goals of its application. The IIoT aims to improve the efficiency of industrial processes by establishing machine-to-machine connections and increasing the degree of automation (Sisinni et al. 2018). IIoT is primarily used in areas such as automation, robotics, logistics, and predictive maintenance planning.

## 3.3 Perspectives

For a complete and accurate discovery and modeling of processes, different perspectives must be considered (Jablonski and Götz 2007; Tiftik et al. 2022; van der Aalst 2016). They offer points of view on different aspects that contribute to a comprehensive understanding. These perspectives are interlinked and cannot be viewed in isolation due to interdependencies. Accordingly, several perspectives should be included in the process model.
Five key perspectives relevant to IIoT processes are identified, each of which contributes to a holistic view of the process (Jablonski and Götz 2007).

**Data Perspective** includes input and output objects of a process, such as documents, files or artifacts and time for provision. The perspective shows how a task is performed and establishes a link between the process and (external) data models (Jablonski and Götz 2007; Dumas et al. 2021). It also allows the effects of data on the process to be analyzed (Tiftik et al. 2022).

As part of the process definition for IIoT processes, it is important to record the connections of machines, measurement sensors, data processing systems and their generated data and forwarding paths to present them in the process model. The network traffic data is therefore a part of the data perspective (Hornsteiner et al. 2024).

**Functional Perspective** defines a process step or subprocess and asks what needs to be done. It serves as the basic unit for the execution of a process and specifies which activity or task is executed, whereby the execution can be carried out either by a person or an automated user (Jablonski and Bussler 1996; Awadid 2017).

**Operational Perspective** encompasses the tools, programs, applications or services that are used within a process step (Schönig et al. 2012; Awadid 2017; Jablonski and Götz 2007). It defines how a process step is executed (Jablonski and Bussler 1996).

**Resource Perspective** enables the examination of process execution with regard to the organizational structure. It considers where the activities are performed and who is responsible for the execution (Awadid 2017). It assigns responsibilities for tasks that are assigned to both human actors and systems (Jablonski and Bussler 1996).

**Control Flow Perspective** focuses on the sequence of process steps without considering time (Jablonski and Götz 2007; Dumas et al. 2021). It specifies the sequence in which activities, functional units and events appear, thus illustrating the workflow and showing the causal dependencies between all modeling elements (Jablonski and Götz 2007; Awadid 2017). The control flow specifications must be adhered to during process execution. These can be modeled using control flow constructs such as sequence, parallel branching, conditional branching and merging (Jablonski and Bussler 1996).

# 4 Related Work

To provide the scientific basis for the design of the artifact presented in this paper, a systematic literature review (SLR) was conducted following Okoli and Schabram (2010). A SLR helps to ensure that potentially relevant information is reviewed as comprehensively as possible and that a selection of the required facts and details can be made. The SLR presented in this paper is based on eight main steps, where first the literature search was planned, then the relevant literature was searched and selected, and finally it was extracted and analyzed. In order to find related work, the following research questions were defined:

**Q1. Are there BPM methodologies and tools specific to process discovery in the IIoT?**

**Q2. How to efficiently discover processes in the IIoT?**

The "Web of Science" metadatabase and the "ACM Digital Library" were used for the literature search. In the "Web of Science", the literature searched was restricted to the databases "Springer Nature", "Elsevier", "Emerald" and "IEEE Xplore" as these are of particular relevance to the community. Only literature published after 2007 and written in English or German is taken into account. The restriction to 2007 is due to the standardisation of BPM at that time.
The following search strings were used to answer Q1 and Q2:

*TI=(("BPM" OR "business process management" OR "process management") AND ("IIoT" OR "Industrial IoT" OR "Industrial Internet of Things" OR "IoT" OR "Internet of Things" OR "Industry 4.0") AND ("discover" OR "discovery" OR "discovering")) OR (AB=("BPM" OR "business process management" OR "process management") AND TI=(("IIoT" OR "Industrial IoT" OR "Industrial Internet of Things" OR "IoT" OR "Internet of Things" OR "Industry 4.0") AND ("discover" OR "discovery" OR "discovering")))*
TI = Title, AB = Abstract

Despite a thorough search, no publications could be found that answered these questions. Therefore, research questions Q3 and Q4 were formulated to address related issues.

**Q3. What connections between BPM and the IIoT have already been explored including process discovery?**

**Q4. Are there already established approaches for general process discovery in (industrial) processes that maybe transferable to the IIoT context?**

To answer Q3, the databases were searched using the search-string below and duplicates were eliminated. In a second step, the content of the literature found was analyzed for IIoT context and relevance to the present paper. Selected papers are discussed in more detail below.

*TI=(("BPM" OR "business process management" OR "process management") AND ("IIoT" OR "Industrial IoT" OR "Industrial Internet of Things" OR "IoT" OR "Internet of Things" OR "Industry 4.0")) OR (TI=("BPM" OR "business process management" OR "process management") AND AB=("IIoT" OR "Industrial IoT" OR "Industrial Internet of Things" OR "IoT" OR "Internet of Things" OR "Industry 4.0"))*

To answer Q4, a second search-string was used to find existing process analysis methods. All duplicates and documents not related to manual process discovery methods were removed.

*TI=("Process discovery methods") OR TI=("Business process discovery")*

In Figure 2, the search process and the steps taken for both research questions are shown using a PRISMA diagram. (Page et al. 2021). In Table 1 the related works and their contents found for the two research questions described above are presented. The following criteria were used to analyze the literature:

- Is there a connection between BPM and IIoT? (*IIoT and BPM included*)
- Is a discovered IIoT process serving as a use case? (*IIoT process used*)
- Is there a description of how the process was discovered? (*Discovery described*)
- Was a manual method used or was the process not discovered by process mining? (*Manual method used*)
- Is it described how processes can be discovered using manual methods? (*Manual method described*)

Our analysis (see Table 1) reveals the absence of specialized manual methods for IIoT process discovery in existing literature, with current efforts predominantly focusing on process mining approaches. While some studies incorporate IIoT processes, they do not detail the methodologies employed for their identification. The key findings from the related literature are elaborated upon below.

The literature review by Giudice (2016) describes the potential impact of IoT on BPM, both inside and outside the enterprise. In particular, these impacts relate to the support of BPM in process optimization and implementation. Although a connection between IoT and BPM is described here, it is the reverse of this paper, as the influence of IoT on BPM is examined. The use of IoT is understood in an industrial context, but not explicitly mentioned. In addition, the literature review does not mention process discovery.

Bazan and Estevez (2022) presents a similar approach to Giudice (2016). They examine the impact of IoT on BPM and emphasize that successful BPM in an organization makes it easier to deal with the challenges of Industry 4.0. Although process discovery is implicitly addressed in this context, it is not explicitly mentioned.
Schönig et al. (2020b) presents an approach for interacting with IIoT data during process execution. This approach enables portable real-time user interfaces to inform process participants about current tasks and thus support efficient task execution. Schönig et al. (2020b) describe an adaptation of process execution for industrial processes using IIoT data, but do not address how the processes used (which are obviously industrial processes) were discovered.
Seiger et al. (2022) Investigate the integration of BPM and IIoT to provide benefits to both research areas. This includes the development of an IIoT system architecture that is integrated with BPM systems in smart factories. The goal is to enable sensor events to interact with existing BPM systems. To achieve this, IIoT technologies will be integrated with BPM technologies along the BPM lifecycle. The focus is on process modeling, automation and process mining. Although process discovery is not a central topic, it is mentioned that process mining techniques can be used to discover processes based on logs and obtain statistics. However, there is no detailed explanation of discovery.
Janiesch et al. (2020), similar to Giudice (2016), discusses the relationship between IoT and BPM. The goal is to highlight the challenges of connecting the previously separate domains of IoT and BPM. Process discovery is also addressed, which is performed using log files and the process data itself. Apart from process mining, no other discovery techniques are described.
Ghose et al. (2007) presents a self-developed method for deriving process models from textual descriptions, which supports the answer to Q4. This approach is based on the assumption that documents serve as a source of information for traditional modeling. Although it is an evolution of document analysis that can accelerate process discovery, this approach does not provide a detailed description of a specific process discovery method. The goal is not to model a process from scratch, but to automatically obtain an initial model that can then be edited by the modeler.
Similar to Ghose et al. (2007), Han et al. (2020) presents an automated business process service that aims to streamline the process discovery phase of BPM projects. Unlike existing approaches that rely heavily on human knowledge to derive hierarchical structural relationships between activities from textual process documentation, the approach uses a neural network. It is designed to automatically find the latent hierarchical structure in the documents. Through process-level language modeling, it can accurately identify process elements and generate BPMN scripts, significantly reducing the time and effort required for process discovery.

| Authors | IIoT and BPM included | IIoT process used | Discovery described | Manual method used | Manual method described |
|---|---|---|---|---|---|
| Houy et al. (2010) | ✓ | ✓ | ✓ | ✗ | ✗ |
| Grefen et al. (2018) | ✓ | ✗ | ✗ | ✗ | ✗ |
| Schönig et al. (2020b) | ✓ | ✓ | ✗ | ✗ | ✗ |
| Schönig et al. (2020a) | ✓ | ✓ | ✗ | ✗ | ✗ |
| Giudice (2016) | ✓ | ✗ | ✗ | ✗ | ✗ |
| Bazan and Estevez (2022) | ✓ | ✓ | ✗ | ✗ | ✗ |
| Seiger et al. (2022) | ✓ | ✓ | ✗ | ✗ | ✗ |
| Schönig et al. (2020) | ✓ | ✓ | ✗ | ✗ | ✗ |
| Janiesch et al. (2020) | ✓ | ✗ | ✗ | ✗ | ✗ |
| Mass et al. (2016) | ✓ | ✓ | ✗ | ✗ | ✗ |
| Schönig et al. (2018) | ✓ | ✗ | ✗ | ✗ | ✗ |
| D'Hondt et al. (2019) | ✓ | ✓ | ✗ | ✗ | ✗ |
| Ghose et al. (2007) | ✗ | ✗ | ✗ | ✗ | ✓ |
| Han et al. (2020) | ✗ | ✗ | ✓ | ✓ | ✓ |

Table 1: Review of thematically relevant literature with regard to the mentioned criteria.

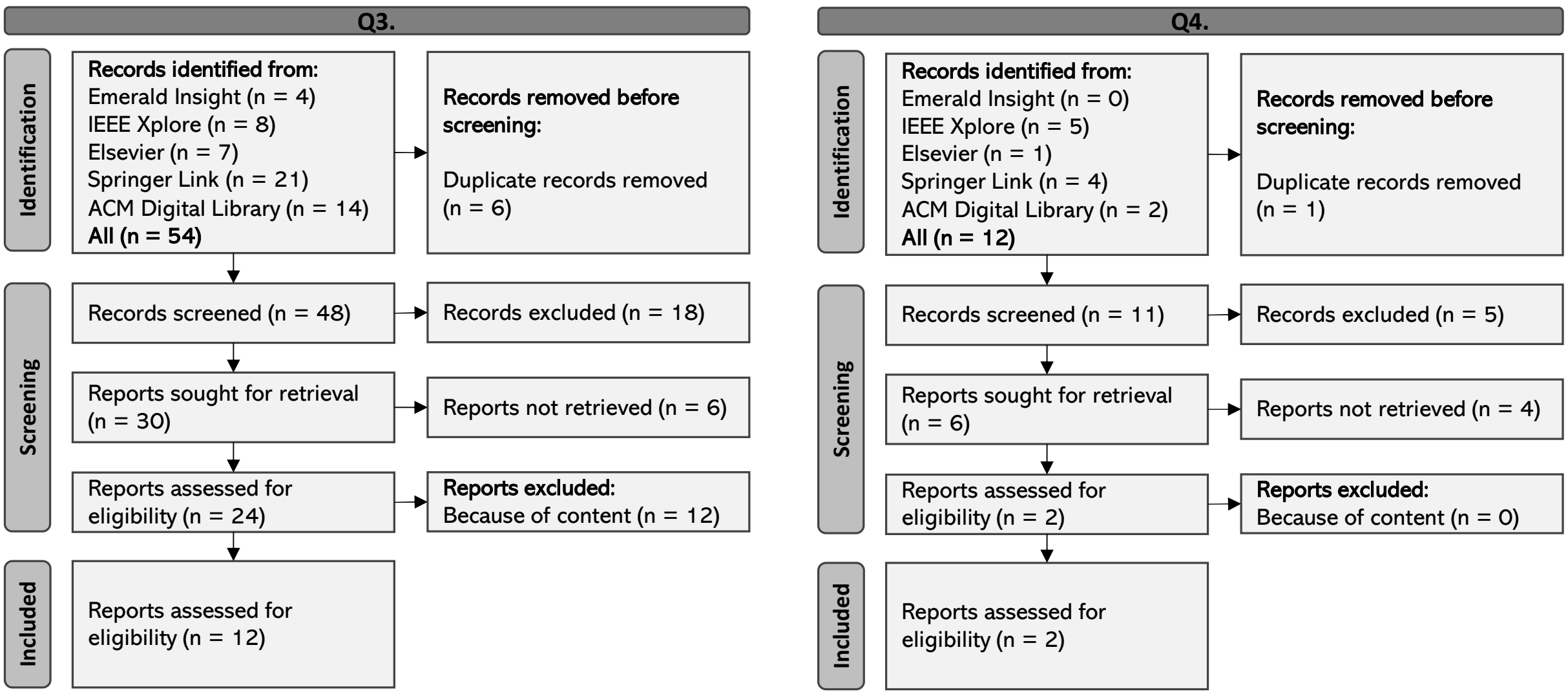

Figure 2: PRISMA-diagram for Q3 and Q4. (Page et al. 2021)

As the literature review revealed that previous literature does not contain manual methods for process discovery in the IIoT context, additional literature was consulted. In particular, Dumas et al. (2021) and Gronau (2017), as these are considered standard literature in process discovery and describe classic methods.

Dumas et al. (2021) provide insights into possible *established approaches to general process discovery*. The methods of document analysis, observation, interviews, and workshops are explained. A description and an example are given for each method. Brief guidelines for conducting interview and workshop-based discovery methods are provided. However, the document analysis and observation methods are covered in less detail. However, the methods presented relate to classic BPM tasks in an office environment, for example, and not to industrial processes.

Gronau (2017) explain methods such as interviews, questionnaires, focus groups (understood as workshops), observations, the inventory method (which includes document analysis) and automatic procedures. The differences between the individual approaches are explained and vary in detail with regard to the implementation of the individual methods. The focus is primarily on the interpersonal component, without any industrial reference.

The search for related work shows that no manual approaches for process discovery in the IIoT context have yet been described. Janiesch et al. (2020) and Seiger et al. (2022) only mention the possibility of process discovery based on log files and data generated during process execution. This identified research gap is closed in this paper by developing structured methods for manual process discovery in an industrial context based on Dumas et al. (2021).

# 5 Guidelines for Manual Process Discovery

This chapter provides an application proposal for the classical discovery methods *document analysis*, *observation*, *interview* and *workshop*. Every subsection is divided into two parts: *(i)* the classical background and fundamentals of the process discovery method and *(ii)* the description of the guideline of the method. The guidelines of the methods follows the given pattern for each method:

1. Specification of the prerequisites for the application of the method
2. Guideline for the method
3. Advantages and disadvantages of the method

To integrate the IIoT, we place an increased focus on the *data perspective* within the guidelines, which depicts the network communication and data utilisation of the processes. The guidelines can also be used for processes without IIoT. In that case, components of the guidelines can be omitted.

The descriptions presented are guidelines that can be applied, but whose suggested procedures do not have to be strictly adhered to. In order to successfully discover a process, the sequence of the discovery can be changed flexibly. At the end of the discovery it is important to ensure that all perspectives from Section 3.3 have been discovered.

Figure 3 shows the requirements that must be fulfilled after a process has been identified. Different discovery methods can also be combined with each other, whereby only individual discovery aspects of the methods can be addressed. An example proposal for combining the methods can be found in Section 7. The individual components of the respective discovery methods are presented here in a possible sequence that can be used by inexperienced process discoverer (PD), for example.

The guidelines are explicitly aimed at people with little or no prior knowledge of process discovery who have no idea how to carry out the classic process discovery methods. People who already have prior knowledge can modify the guidelines or pick up on individual aspects if necessary.

We recommend drawing up a corresponding process map and identifying relevant processes within this map before collecting processes. The process to be discovered should be considered in the context of the previous and subsequent processes. In addition, the objective and purpose of the process discovery should be defined prior to discovery in order to adjust the focus and level of detail accordingly. The discovery can be carried out top-down, depending on the desired level of detail.

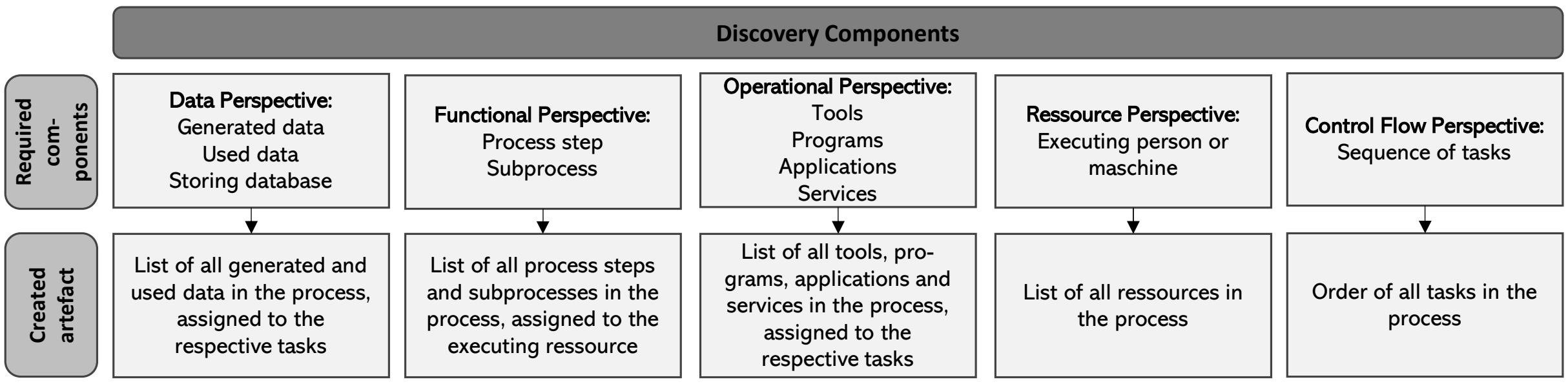

Figure 3: These process components must have been collected at the end of the process discovery.

## 5.1 Document Analysis

### 5.1.1 Classical Document Analysis

Existing process descriptions such as internal policies, organizational charts, manuals, models or instructions can be used for process discovery (Dumas et al. 2021). Similarly, information from business distribution plans, regulations, budget, economic, personnel, and staffing plans, directives, job descriptions, requirement documents, design documents, order statements or meeting protocols can be obtained by the PD (Federal Ministry of the Interior and Community 2012). The collected documents, containing information about the current state of the process, are reviewed and process-relevant information is extracted (Federal Ministry of the Interior and Community 2012). Based on this information, a model of the process can be created. During document analysis, the following information can be extracted for process discovery (Federal Ministry of the Interior and Community 2012): Task structures, areas, and carriers; organizational structures and process workflow; specifications and conditions within the process; decision points; time required; linked and used documents in IT systems. There is no information on the approach or techniques for extracting information from the documents and converting it into a process model. Documents are usually not process-oriented and therefore not uniformly granular. This leads to different levels of detail in the various documents, making direct derivation of processes from the documents usually not possible (Dumas et al. 2021). Furthermore, the documents used may be outdated or incorrect and do not guarantee the correctness of the information contained therein as they refer to past states (Dumas et al. 2021; Hansen et al. 2019). However, documents are free from personal evaluations and thus provide an objective basis for familiarizing oneself with a process. They serve as a starting point for process discovery (Dumas et al. 2021; Hansen et al. 2019; Federal Ministry of the Interior and Community 2012).

### 5.1.2 Document Analysis in IIoT

For IIoT process document analysis, specific requirements and assumptions should be met:

- Sufficient documentation must be available that the process can be derived from these materials.

- The documents must be available in a format understandable to the PD.
- The documents must be up-to-date, error-free and detailed.
- To discover IIoT-processes the connectivity of the machines, sensing sensors, produced data, data processing systems and workflow must be documented.

**Preparation** The preparation of the document analysis starts with the collection of all process-relevant documents. These include for example business distribution plans, business regulations, budget, economic, personnel, and staffing plans, instructions, job descriptions, requirement and design documents, order declarations, machine manuals and operating instructions, specification documents for involved individuals, site plans of the production facility and other documents related to the process or involved entities.
Sort the documents according to their level of detail and check short and structured documents first, followed by more extensive documents. Finally, focus on machine-specific documents due to their specificity. This top-down approach provides a clear overview before delving into details systematically.

**Discover Resources** In this step, the documents are reviewed from the *resource perspective*. All documents are examined to determine which executing instances, i.e. people and machines (short: resources), are involved in the process execution. Each document is worked through individually, and notes are taken as soon as a machine or a role of a person performing a specific activity is mentioned. Person roles are often terms describing activities, i.e. "truck driver". Machines are usually identified by their own names i.e. "welding machine". If explicit names for machines or person roles are not provided in the documents, the PD should assign appropriate designations. It is crucial that these designations are used consistently to avoid confusion, duplicates, or other errors.

**Discover Tasks** As soon as all resources have been identified and noted, we recommend recording the associated tasks, the *functional perspective*, and assigning them directly to the discovered resources. All documents should be reviewed and every activity and task (short: task) mentioned should be noted. It should also be noted which resource performs the task that this can be assigned directly during modelling later. A task is one step of work in the process. It is important to capture all tasks that can be performed by a resource. This applies regardless of whether the tasks appear to be related, occur only in special cases, or have other peculiarities. If a task can be performed by different resources, the task is noted for each participating resource, along with an indication which other resources can also perform this task. This will be taken into account later in the modeling by creating an exclusive decision node, which ensures that the task is only executed once. It is also possible to discover the tasks independently of the previous resource acquisition. Nevertheless, it should be ensured that the tasks can be assigned to the correct resource later.

**Discover Tools** The *operational perspective* is captured by discovering the he tools, programs, applications, services, and software components (short: tools) used to perform a task. For each task, it is determined how it is implemented in practice. All available documents must be reviewed and searched for the mentioned tools and the task where they are used. A direct assignment of the tools to the task is recommended directly when the tool is discovered. For example, if the ERP system is used, this is noted for the corresponding task. It is also possible to detect the tools independently of the previous task detection. Nevertheless, it should be ensured that the tools can be assigned to the correct task later.

**Discover Data** The data which is produced or used in each task covers the *data perspective*. The documents should be read through chronologically and any data element mentioned should be noted. For each data element noted, the task in which it is produced and used should be noted. If a data element is produced in a task, details such as the type of sensor used, data capture intervals, data format, and destination are useful to record. If a data element is used in a task, the purpose, format and source of the data are possibly interesting. In both cases, the storage location of the data, typically a database, is also indicated. In order to manage the collected information clearly, it is recommended to use tables that record the aforementioned properties as columns and the data elements as rows.

**Discover Sequence** The sequence of individual tasks is extracted from the documents to represent the *control flow perspective*. All documents are read through, examining which task the process starts with and which subsequent tasks follow. It is advisable to identify the starting point of the process and then sequentially determine the next step. If there are decision points or intermediate events within the process, these should also be recorded. This may include information such as the duration of certain process steps, the conditions under which certain tasks are executed and identifying which tasks sensors produce data.

**Validation** Whether the process has been accurately captured cannot be immediately verified. There is the possibility of conducting the entire document analysis again and recreating the model independently of the initial capture. An alternative method for validating the model is to choose another discovery method and execute it. Subsequently, the two process models can be compared. If discrepancies are found between the two models, this indicates that at least one of the captures was flawed or incomplete. In such a case, it is necessary to review the documents again to identify and rectify any errors and incompleteness.

In order to collect the individual perspectives, the documents can be run through in several iterations for one perspective at a time or all perspectives can be run through within one iteration. For the sake of clarity, we recommend proceeding in several iterations and identifying the resources first. Then collect the tasks and assign them to the corresponding resources. Collect the data and tools at the same time and assign them directly to the tasks. Finally, the tasks can then be put in the correct order using the documents. Table 2 encapsulates the advantages and disadvantages of document analysis.

## 5.2 Observation

### 5.2.1 Classical Observation

During observation, the PD observes the execution of the process and can directly capture the current process activities (Dumas et al. 2021; Gronau 2017). In business processes, the PD can take an active role (e.g. the perspective of the customer) or a passive role (e.g. the perspective of the employee) to conduct overt or covert observations (Dumas et al. 2021; Gronau 2017). The PD documents their observations and subsequently translates them into a process model. Similar to document analysis, the literature does not provide specific techniques or instructions for conducting observations. Since observation is passive, no workflows or processes are interrupted. There are no additional costs due to work interruptions or production stops during process discovery (Gronau 2017; Hansen et al. 2019; Dumas et al. 2021). Furthermore, observation enables the PD to gain a better understanding of process boundaries and reduces reliance on information from documents or process participants (Dumas et al. 2021). Observation allows for an objective view of the process from the perspective of the PD and is not subject to the subjective influences of process participants. However, it should be noted that employees may change their behavior if they know they are being observed, and may work in a more structured or faster manner than usual (Dumas et al. 2021; Hansen et al. 2019). Additionally, observation requires that the PD has access to the process, which may require access to remote facilities or locations (Dumas et al. 2021). It may be temporally or spatially impossible to continuously observe a process. In such cases, the PD may need to observe the process multiple times to ensure that all possible process paths have been observed at least once.

| Advantages | Disadvantages |
|---|---|
| • No adverse effects on ongoing production processes<br>• No production outages or hindrances<br>• No additional resource costs for personnel<br>• Detailed documents facilitate comprehensive process depiction<br>• Continuous availability of documents for reference | • Significant time investment for the PD<br>• Increased analysis time with more documents<br>• Possibility of not fully understanding technical machine descriptions<br>• Need for involvement of additional individuals for technical understanding<br>• Inaccurate or outdated process models due to erroneous or incomplete information<br>• Lack of detailed documents on crucial aspects like data and machine communication<br>• No extensive documentation available in IIoT processes |

Table 2: Advantages and disadvantages of conducting document analysis in IIoT processes.

#### 5.2.2 Observation in IIoT

For IIoT process observation, specific requirements and assumptions should be met:

- The process is located within one or more production facilities where the PD has access.
- The PD must be physically present on-site at the production facility.
- The activities performed by machines must be observable that the PD can see them.

It is recommended to model the process during observation. This allows the process to be captured based on the sequence of execution, enabling the components to be recorded directly in the correct order. Thus, the guiding perspective for observation is the *control flow perspective*.
The PD tracks the manufacturing of a particular component from the beginning to the end of the process. For example, if a piece of wire is clamped into a machine at the beginning, the PD follows this piece of wire until the end of the process, where a complete final product is produced. The PD starts at the beginning of the process and records the trigger of the process. Then, they track the first process step. If there are multiple ways the process can start, one possible starting event is chosen, while all other possible starting events are noted. After the observation is completed, the process is observed from another starting event until the first common task reached after all starting paths.

**Discover Tasks** For each process step, the PD observe what happens in the process step to cover the *functional perspective*. A process step in the observation is an executed activity and is thus differentiated from the next or previous step. It should be possible to record the observed activities as compactly as possible. The level of detail of a task depends on the previously defined granularity of the process.

**Discover Resources and Tools** At the same time as observing the task, the PD can make a direct note of who performs this process step (which resource) and whether it is apparent which tools are used for execution. This covers the *resource* and *operational perspectives*. The resources and tools can also be discovered independently of each other and of the tasks. However, it is important that they are allocated correctly. We therefore recommend assigning a task directly to the executing resource and noting the tools used. The executing resource can be directly modeled as a lane. If it already exists, the step is skipped. The observed task is modeled as a task and assigned to the corresponding resource. Then, any tools used are linked to the task. Once all components are captured, the PD moves on to the next process step and repeats the process.

**Discover Sensors for the Data Perspective** The collection of data perspective is not possible in the observation. Nevertheless, sensors that are sighted during the process can be included in the model. This allows the model to show where data is generated in the process.

**Handling Branches** Once the PD recognizes that the process branches, it is advisable to follow one path at a time and not try to capture multiple paths simultaneously, because the size and complexity of the process can lead to confusion. The model could become faulty or incomplete if the PD forgets parts of the process or neglects a path. To maintain clarity, the following approach is recommended when encountering a process branch:

1. The PD notes in a separate document the task where the process branched and the number of possible additional paths at this point.
2. The PD choose one of these paths and continue through the process.
3. Once the PD reach the end of the selected path, they return to the marked point and choose the next path.
4. The process is fully traversed again, and this process is repeated until all noted branches are addressed.

If there are further branches of the process in a path, the task and the number of paths are noted. The processing of the noted branching points is done chronologically according to the notation, with initially noted branches being processed first. This ensures that all branches are addressed, and the process is modeled along the control flow.

**Handling Overlooking of Paths** The overlooking of paths in forward observation occurs when paths are rarely executed. If a path is not activated during observation, it cannot be observed and therefore cannot be captured in the model. However, backward observation does not guarantee the capture of rarely executed paths. To reduce the likelihood of overlooking paths, the following measures can be taken:

- *Multiple iterations:* Observation can be conducted multiple times, both forward and backward. Each additional iteration increases the chance of capturing rarely executed paths.
- *Increased observation duration:* To ensure that even rare paths are captured, the observation duration at each process branch can be extended. This allows for the discovery of unusual or less frequent process flows.

**Validation** The accuracy of the observation process discovery can be validated. To do this, the observation is repeated from the end of the process. This ensures that paths or tasks that may have been overlooked in the forward observation are discovered in the backward observation. The observation begins usually at the finished product. Each previous task is considered and compared with the process model. Special attention is paid to process path mergers. This occurs when a task has multiple possible preceding tasks, resulting in multiple process paths leading to this task. The observed task is noted together with the number of triggering options, as with forward observation. Subsequently, one of these possibilities is selected, and the observation is conducted backward until the beginning of the process is reached. If additional path mergers occur during this backward observation, they are also re-noted. All mergers should be processed chronologically until all have been traversed. The goal of backward observation is to uncover paths that were overlooked during forward observation. For example, if only three possible paths were identified at a process branch, when there are actually four possible paths, the backward observation might uncover four converged paths. Once all these paths have been traversed backward, a path that is not included in the current model can be identified and added.

**Using Observation for IIoT Processes** Although it is not possible to collect the *data perspective*, observation is suitable for recording IIoT processes. This is because prior observation helps the PD to understand the process and provides initial insights into the process, which the PD can use to develop questions and initial models based on their prior knowledge using other discovery methods, e.g. in an interview. This is shown in Section 7.

Table 3 encapsulates the advantages and disadvantages of observation in IIoT processes.

| Advantages | Disadvantages |
|---|---|
| • Allows an objective view of the process<br>• Overview of all process components<br>• Observation can be conducted within a spatially limited framework, such as a production facility<br>• Does not disrupt process flows<br>• No additional resources or costs required<br>• Process flow is quickly evident to the PD | • Not all process operations are visible, particularly those within machines<br>• Data usage, such as collected data, network traffic, and machine communication, remains hidden<br>• Some process steps may not be observable due to machine placement or internal processes<br>• Risk of omitting rarely executed process paths, rendering the model incomplete<br>• Collection of waiting times can slow down process discovery speed and consume time |

Table 3: Advantages and disadvantages of conducting observations in IIoT processes.

## 5.3 Interview

### 5.3.1 Classical Interview

In the interview-based method, a subject matter expert (process owner) involved in the process is personally interviewed (Dumas et al. 2021; Gronau 2017). If multiple process owners are involved in the process, multiple different interviews must be conducted. Dumas et al. (2021) propose two strategies for conducting interviews: *(i)* backward and *(ii)* forward process discovery. In *(i)* the interview starts at the end of the process and works backwards through the process, asking "What happens before?" and continues until the trigger of the process is reached. *(ii)* operates in reverse, collecting data on the process from the beginning to the end state of the process. Furthermore, Dumas et al. (2021) recommend choosing between a structured or open interview approach. Structured interviews involve predefined questions and hypotheses that need to be confirmed or corrected. This ensures that specific questions of the PD are answered. However, important information may be overlooked in a structured interview as it may not be included in the predefined questions. Additionally, exceptional cases may remain unaddressed. Open interviews allow the process owner to present important aspects of the process from their perspective, potentially revealing information that would have remained hidden in a structured interview. On the other hand, the process process owner may omit important information, thus keeping it hidden from the PD (Dumas et al. 2021).

To conduct process interviews, the Bundesverwaltungsamt (2013) has created a *"Guideline for the Collection of Business Processes"*. To prepare the interview, existing documents should be reviewed and a structure of the process should be prepared based on their information. Subsequently, this information must be verified during the interview. They provide tips for interpersonal aspects that facilitate conducting interviews. No further guidance on structured interview conduct is provided. Gronau (2017) point out general principles for conducting interviews:

- Interview questions must be formulated and asked precisely.
- The PD must be aware of their role as an interviewer. Their interpersonal skills influence the

success of the interview and the relationship with their interviewees.

- The information obtained must be carefully documented.
- The interview questions must be tailored to the expertise of the process owner.

The process model is created from the interview notes or parallel during the interview (Dumas et al. 2021). Overall, conducting interviews requires significant time and resource investment. When multiple process owners are involved in a process, contradictions may arise that the PD must clarify. This requires verifying the correct process flow, leading to increased time investment (Dumas et al. 2021). Qualified subject matter process owners familiar with the process details are required. During interviews, subject matter process owners may not be able to perform their regular duties, potentially impacting operations (Krallmann et al. 2002).

### 5.3.2 Interview in IIoT

For IIoT process interviews, specific requirements and assumptions should be met:

- One or more persons (referred as the process owner) who are familiar with the process or involved must be available.
- The process owner(s) needs to understand the process or the subprocess they are responsible for in detail and can describe it.
- The process owner(s) possess a comprehensive understanding of the IIoT characteristics of the process and can identify and explain them.

One or more interviews can be conducted for the interview-based discovery. This depends on the complexity and size of the process and the personal preferences of the PD and the process owner. We recommend conducting a separate validation interview after the discovery, which takes place on a separate date, to check the correctness of the model once it has been fully modelled.

**Interviews with Multiple Process Owners** If processes are large and complex (especially IIoT-processes) they can be divided into different sections. Each section typically has different process owners. To conduct a process discovery through interviews, the PD should invite all process owners to separate interviews. These interviews can follow or use parts of the guideline described below with each invited participant separately. Once all sections of the process are discovered, the PD must assemble them in the correct sequence. If there are interactions or cross-connections between the interviews or if process owners contradict each other, it is advisable to talk to the process owners involved together to clarify the affected process component.

**Preparation through Questionnaire** At the outset, the PD should provide the process owner with a preliminary questionnaire regarding the process. The questionnaire should include questions covering all process perspectives. It is important to use terms in the questionnaire that are familiar to individuals without a background in process modeling. Possible questions for the questionnaire could include:

- What is the main goal or output of the process? What is produced, and what is the outcome of the process?
- Are there specific triggers or conditions starting the process?

- Which individuals, machines and IT systems are involved in the process? (*Resource perspective*)
- What is the flow of the process? Please describe chronologically what happens in the process. (*Functional and control flow perspective*)
- Are supporting software or other tools used to perform tasks? If yes, please name the tasks where such support is provided and describe the type of software, tools, or other aids involved. (*Operational perspective*)
- Are there specific procedures or measures that are executed in case of an error occurrence?
- What happens to unusable or defective parts identified during the process? Are there mechanisms for monitoring and segregating such parts? Please explain these.
- Which sensors are present within the process? Where in the process are these sensors used (at which tasks)? What data do these sensors each capture and produce? Please describe this briefly. (*Data perspective*)
- Where are the captured data sent? Where does data processing take place? (*Data perspective*)
- Which machines within the process are interconnected? How do these machines communicate with each other? What information or data do the machines exchange?
- Is there a data network within the process? If yes, please describe and sketch it briefly. (*Data perspective*)

Based on the answers, the PD can create an initial process model, considering the various perspectives. The questionnaire allows minimizing the duration of the subsequent interview and gives the PD the opportunity to familiarize themselves with the process before the interview. The PD can already prepare more detailed questions for the first interview, considering the following topics:

- Have any uncertainties or contradictions arisen due to the answers from the first questionnaire?
- Is the process model complete or are there paths that do not have a defined end?
- Does the sequence of tasks in the process make sense?

**Starting the Interview** We recommend dividing the interview into several parts to allow the PD and the process owner to take breaks. However, it is also possible to conduct only one interview where only parts of the proposed methods are used.

The PD invites the process owner to the interview. Two forms of interview can be conducted: a closed interview with prepared questions that already explicitly address specific content. This is particularly recommended if a questionnaire has been sent out in advance. During the interview, the PD asks the process owner the prepared questions and notes down the answers. Otherwise, in an open interview, the process owner can be explicitly asked to talk about the process on their own initiative using open questions. This gives the process owner the opportunity to address additional comments and ensures that the PD receives information that may not have been explicitly asked for. Examples of conducting interviews using open and closed interviews can be found in Dumas et al. (2021). Further literature on conducting interviews is provided by Taherdoost (2022), Turner III (2010), Gubrium et al. (2012) and DiCicco-Bloom and Crabtree (2006).

**Check Understanding** To check if the PD has understood the answers and process descriptions of the process owner correctly, we recommend carrying out the following step: The PD goes through their notes with the process owner and asks them about each process step, confirming whether the representation and understanding of the process flow are correct to avoid potential errors in modeling. If there are any errors or incompleteness, the process owner correct them. Alternatively, the PD can integrate the information received during the interview into the existing process model and expand it or, if no model is available yet, model it. If a model already exists, the PD goes through the process model with the process owner and explains the modelled steps. Here again, the process owner should point out any errors or incompleteness and make corrections. It is important to note that the PD should never assume that the process owner is familiar with the modeling language. Therefore, the PD should explain each modeled process step by describing what happens at that point according to the model. If a correction or addition is necessary, the process owner should communicate this verbally. The PD can either note this information and integrate it into the model later or make the changes directly in the model.

After going through the notes or model once, the PD should ask the process owner if there are any additional information or process components that have not been captured yet (e.g. a process branching that triggers another, previously unrecorded path). If such information is available, it should be integrated into the notes or the model. If desired, a pause can be taken between the interviews at this point.

Depending on the duration of the first interview and the extent of the model changes, it may be useful to end the interview at this point and conduct a second part in a separate session. This allows the PD to integrate new insights from their notes into the model without causing unnecessary waiting time for the process owner. Updating the model can take several hours depending on the extent of the notes and the complexity of the process. Additionally, the concentration of the participants decreases after a certain duration.

**Detail Contents** In a second interview or as a second step, the process can be detailed and process exceptions identified. It can be assumed that the first model does not have a high level of detail yet and that none of the rarely executed paths have been captured. These details should be worked out in the second interview. This includes identifying subtasks, defining decision rules at relevant branches, integrating sensors at suitable locations in the model for data collection, documenting the data forwarding and usage at the respective goals, and verifying that all possible process paths are mapped in the model, regardless of their rarity of execution. To collect this information in a structured manner, the PD can question the process owner based on the revised model from the first interview. It is recommended to go through the following questions for each task to capture all necessary information:

- Does the task divide into further subtasks? If yes, into which ones?
- Are data collected at this point in the process? If yes, which ones, why are they collected, where are they sent (to which other process task), and where are they stored (at which storage location)?
- Are there additional paths that can follow this task that have not been mapped so far?

The PD may further divide tasks and create multiple tasks or model subprocesses if necessary. Data are attached or referenced as data objects to the respective tasks, and connections are documented. Once all tasks have been traversed, all branches are traversed chronologically, and for each path, it is determined what conditions must be met for the path to be executed, the interview ends. The PD can add the information into the model now. In order to pursue the top-down approach, the interview can also be conducted several times in order to discover the process more and more granularly until the desired level of detail of the process is achieved.

**Area Expert** If the process owner (contrary to the assumption) does not have sufficient detailed knowledge of the process to answer the questions and describe the process at a certain point, an area expert must be involved. This person is also familiar with the process but has specific expertise in a particular area. The area expert is called in for a discussion and supports the process owner where they lack detailed knowledge of the process. The area expert is explicitly involved only for this specific section of the process.

**Validation** We recommend using a validation interview where the PD and the process owner again go through the extended process model to check for correctness and completeness. The process should traversed chronologically and at each task it is checked whether the task is correct, the associated data is correct, the executing resource is correctly assigned, and the task order is correct. It should be verified whether the tasks before and after are actually executed at this point in the process flow. When the process reaches a branching point, its paths and the conditions required for them are checked. Subsequently, the cooperation between the individual machines and roles is examined, with each connection being verified to ensure it actually exists and connects the correct resources. If the process owner has no corrections and confirms that the model is complete and correct, the process discovery is completed. If errors are found by the process owner, they should be corrected after the interview. The interview can be repeated until the process owner confirms the model as correct and complete. However, the model can also be terminated after the corrections of the validation interview or terminated after a certain number of validation steps. The PD decides when the model and the process discovery are complete. Table 4 encapsulates the advantages and disadvantages of interviews in IIoT processes.

## 5.4 Workshop

### 5.4.1 Classical Workshop

The workshop-based method is similar to the interview method, but involves several process experts simultaneously (Dumas et al. 2021; Jadhav 2011). When selecting participants, it should be considered that small, hierarchically homogeneous groups work more efficiently than large groups or groups with hierarchical heterogeneity. It is not recommended to include executives and their employees in the same workshop (Richerzhagen 2015; Dumas et al. 2021). Technical staff that is involved in the process indirectly, for example, through the management of supporting systems (e.g. ERP systems) should also be involved. However, the maximum number of twelve experts in a workshop should not be exceeded (Dumas et al. 2021). Jadhav (2011) recommends conducting review workshops, where subject matter experts must approve the modeled process afterwards to ensure that the process is consistent and correct. Dumas et al. (2021) suggest that a moderator moderate the workshop and should coordinate the contributions of the participants and equally involve the process experts in the workshop. Involving a moderator allows the process discoverer to design a model during the workshop.
In addition, a process analyst can be included as a third person to note relevant statements made by the participants, which may need to be followed up on later that no interruptions of the discussion for questions is needed. This approach is particularly suitable for complex processes involving many experts (Dumas et al. 2021). If there are different opinions within the group regarding the content or presentation of the process, it is the moderator's task to steer the group discussion that ultimately the correct process can be identified (Bundesverwaltungsamt 2013).

| Advantages | Disadvantages |
|---|---|
| • Provides up-to-date information from the process owner<br>• Content-related queries and misunderstandings can be clarified during collection<br>• Ensures all relevant information can be collected, including details potentially hidden during observation or not mentioned in documents<br>• Allows querying of all possible process paths, data, and network information<br>• Provides extensive means for gathering IIoT-related information about a process<br>• Increases likelihood of having a complete and accurate model | • Time- and cost-intensive<br>• Disruption of process flows as the process owner cannot carry out usual activities during interviews, potentially incurring higher costs<br>• Reliance on the assumption that the process owner knows the process comprehensively or receives assistance from area experts<br>• Possibility that the process owner may not provide correct and complete information, leading to inaccuracies in the model<br>• Risk of the process owner describing an idealized future process rather than the actual current process, leading to models that do not correspond to reality |

Table 4: Advantages and disadvantages of conducting interviews in IIoT processes.

Since a detailed model cannot be created in one session due to complexity, multiple sessions are required (Dumas et al. 2021). The first session is used to establish the context, communicate the workshop's goals, and identify the participants' expectations (Jadhav 2011; Dumas et al. 2021). In addition, the scope of the organisation to be considered must be defined, whereby the use of process maps, if available, is appropriate (Richerzhagen 2015).

### 5.4.2 Workshop in IIoT

For IIoT process workshops, specific requirements and assumptions should be met:

- At least two or more people need to be involved in the process (referred to as process participants).
- The process participants should possess detailed knowledge about the process or its components, including an extensive understanding of the IIoT characteristics of the process.
- The process participants must be available simultaneously (a common schedule must be feasible).

A process discovery through workshops should involves the preparation and at least one or more workshops. The workshop-based discovery can be carried out in one or in several workshops. The PD can incorporate the information into the process model between the workshops. Otherwise, the model can be created in parallel to the workshop or afterwards. For the workshop, we also recommend analysing the individual perspectives and pursuing a top-down approach. For tips on moderation and interaction within a workshop, see the literature by Lauttamäki (2014), Linds and Gee (2023), Pavelin et al. (2014), Jolles (2017) and Storvang et al. (2018).

**Preparation** To conduct a process discovery through a workshop, the PD has the opportunity to decide whether to moderate the workshop and gather information simultaneously or delegate one of these tasks to another person. Additionally, as suggested by Dumas et al. (2021), an analyst may be involved. The planning of the workshop begins with the selection of relevant participants. The PD needs to determine which participants have a detailed understanding of the process. It is important to ensure that at least one participant from each process area attends the workshop to ensure that the entire process is adequately represented by different participants. Participants can include employees who perform direct tasks in the process, as well as those who are indirectly involved in the process, i.e. machine operators or developers of the programs executed in the machines. The PD schedules the dates for the workshop and invites the selected participants, taking care to find dates when all invited participants are available. For the execution of the workshops, it is advisable to follow the suggested approach of Camunda BPM by Richerzhagen (2015).

**Starting the Workshop** At the beginning, the PD should introduce the process being elicited and briefly explain the planned approach for the discovery. Then, all participants should introduce themselves and explain their involvement in the process by describing their roles and providing rough descriptions of their tasks.

**Set up a Process Profile** We recommend starting the workshop by creating a process profile to get a overview of the process. To do this, the participants are asked to list the roles involved in the process (*resource perspective*) and jointly describe a rough process flow (*control flow perspective*). When describing the rough process flow, the participant responsible for each area should explain what happens in that section of the process. The moderator should ensure that all participants contribute equally to the discussion and asks targeted questions. The PD records the information gathered and may directly capture it as a model, while the analyst takes note of additional information that may require further investigation through follow-up questions.

Once all necessary information has been gathered, the analyst can raise any follow-up questions. These questions should be answered by the participants, allowing the PD to directly capture the information. Then, the PD presents their findings and shares their understanding of the process flow, the roles involved and process outcomes with the participants. Participants are encouraged to make any necessary corrections and provide additional comments. This discussion is also guided by the moderator. This part of the workshop is the ideal place to take a break that the PD then can document the results in the form of a process model.

**Working in Groups** Depending on the group size, the process discovery can be done collectively with the moderator in the entire group or, for a large number of participants, in separate groups working independently. Working in small groups is advisable if the moderator deems the overall group to be too large. When working in small groups, the process should be divided into several sections, with each small group assigned a specific section of the process. Each small group should include participants who hold roles in that section of the process. If a role appears in multiple sections of the process, the participant is assigned to the section they consider to be the most complex or the group work is not carried out at the same time to allow the participant to be present in both groups. Once the work in the small groups is completed, they present their results. This is done in the order of the process flow. After all small groups have presented their results, further additions and corrections are discussed in the entire group. Participants, who were required in several stages of the process, were asked to contribute to this discussion. The PD notes the comments for each section. If the workshop is conducted with all participants simultaneously, the work is similar to that in small groups.

**Discovering Tasks, Tools and Data** We recommend to focus on the *functional*, *operational* and *data perspectives* after discovering the process profile. The PD should present the current process model if this is already existing. Subsequently, the refinement of the process and the identification of operational support are carried out.
The hole group or each small group discusses, based on the current process model or information, the refinements needed in each task and whether all tasks for each role in that section are mentioned (*functional perspective*). The additions and refinements should be documented in writing.
For each task, the software and other support tools used should be noted chronologically (*operational perspective*).
A list can be created that enumerates all data generated or used during the process to focus the *data perspective*. It is then recorded for each data record which task generated it, what type of information it contains and where it is used in the process. If the data originates from another process or is sent to another process, this must also be documented. Together with the participants, the PD then assigns the respective data to the producing and consuming tasks, thus establishing the connections between the machines. Finally, the storage location for all data objects (usually the respective database) is determined and noted. The process steps could be detailed and refined, and the support tools used are added. Subsequently, the PD should integrate the new information into the process model. The PD can directly represent this in the modeling tool and make the process model visible to all participants, for example, by using a canvas projection. The order and level of detail in which the various perspectives are recorded can be customised.
The moderator has to ensure that each discussion is continued until the participants agree on the correctness of the discussion component. It is important to ensure that participants are not persuaded to make a statement or agree due to hierarchy, as this could jeopardize the correctness of the process. Higher-ranking employees should not have decisive authority.

**Complete Workflow** To complete the model all decision nodes in the model should be reviewed. This can be done during the creation of the process profile, separately at the end of the workshop or in a separate workshop. When this discovery should take place can be made dependent on the size and complexity of the process and the number of decision nodes. At each decision node, discussions are held regarding the conditions triggering the paths, and the triggering data values are noted. Once all decision nodes have been examined, the analyst asks questions about the recorded information. Once these questions are answered, the workshop is concluded, and the PD can finalize the process model.

**Validation** In the final workshop, the PD presents the completed model and explains the entire process flow to the participants. If all participants agree with the presented process model, the workshop, and thus the process discovery, is considered complete. If there are any improvements, corrections, or additions, these are noted and added into the model. The final workshop can be repeated until all participants agree with the process model or, as with the interview, can be terminated after a certain number of runs or other cancellation criteria. Table 5 encapsulates the advantages and disadvantages of workshops in IIoT processes.

| Advantages | Disadvantages |
|---|---|
| • Opportunity to involve multiple individuals simultaneously<br>• Direct resolution of contradictions<br>• Examination of the process from various perspectives<br>• Availability of experts for different process areas with deep understanding of IIoT-relevant information<br>• Increased likelihood of collecting a correct model compared to interviewing a single process owner<br>• Ability to collect data and data flows in the process, mapping network and connectivity of involved machines | • Higher personnel costs due to participants unable to carry out regular activities during workshops<br>• Scheduling and time investment become more complex and time-consuming with increasing participants<br>• Disruption of process flows as the process owner cannot carry out usual activities during interviews, potentially incurring higher costs<br>• Possibility that none of the participants are informed about certain technical aspects of the process |

Table 5: Advantages and disadvantages of conducting workshops in IIoT processes.

## 6 Combination Opportunity Example

As mentioned in Section 5, it is possible to combine the components of the guidelines with each other as required. This makes it possible to avoid the disadvantages of the individual methods or to give them less weight. In this chapter, we present a possible combination and explain the reasons and objectives of such a combination. By combining the methods, we focus to create a comprehensive, accurate and verifiable model that contains all IIoT-relevant information. We want to ensure that the time and cost investment are kept to a minimum without compromising the completeness, accuracy, and verifiability of the process model.

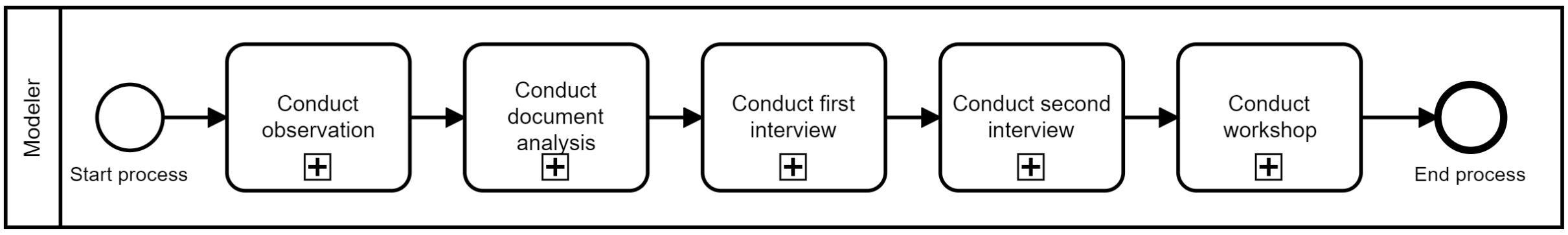

Figure 4: Combination of the individual methods into one process discovery.

This combined method example starts with observing the process. This is followed by a brief document analysis and an interview. Subsequently, another interview and a workshop are conducted for validation purposes, as shown in Figure 4.

The observation is intended to provide the PD with an initial insight into the process and allows for the design of a rough model containing the involved resources and their visibly executed activities. Through observation, aspects of the *resource*, *control flow*, and *functional perspectives* are partially covered. To conduct the observation, the PD follows the instructions outlined in Section 5.2.2 and pass through the process, noting the sequence of performed tasks and the corresponding executing resources. Process branches are handled as explained. The result of the observation is an initial process model where the resources are modeled as lanes and their executed tasks are modeled as tasks in the correct sequence.

Next, the PD proceeds to document analysis and reviews the existing documents for the following information in the given sequence:

1. Additional tasks and paths that were not captured during observation. (Completion of the *control flow* and *functional perspective*.)
2. Software used and supporting tools related to each task. (Discovering the *operational perspective*.)
3. Produced and used data related to each task. (Discovering the *data flow perspective*.)
4. Connections between the involved machines. (Discovering the connectivity.)

To obtain this information, the PD iterates through the documents multiple times. In the first iteration, the focus is on the *control flow* and *functional perspective*. In the second iteration, the *operational* and *data perspectives* are captured. In the third iteration, information about machine connectivity is gathered. Subsequently, the PD also notes any uncertainties and questions arising from the existing model. For the document analysis, the PD should refer to Section 5.1.2. If the PD finds contradictory information in the documents compared to what was collected during the observation, this information from the documents should be ignored as its accuracy is not guaranteed, unlike the observation. Upon completion of the document analysis, the PD already possesses a comprehensive and detailed process model. The correctness of the model has been partially ensured through observation. So far, only costs and time commitments for the PD have been incurred.

In the third step, the process owner is invited to an initial interview session. As mentioned in Section 5.3.2, it may be necessary to interview multiple process owners. The following formulations refer to a single process owner, but can also be optionally conducted with multiple process owners for different process areas. A detailed questionnaire is not sent in advance to avoid further increasing the time commitments of the process owner. Instead, the process owner is simply informed about the process that will be discovered and provided with the questions that the PD has noted for preparation.

In the interview, the PD proceeds according to the instructions in Section 5.3.2. The PD asks the previ-ously noted questions and records the answers from the process owner. The process owner then has the opportunity to contribute additional relevant aspects. Afterwards, the PD presents the current process model. The modeled process is then worked through chronologically, and the following questions are clarified for each task:

- Does the task break down into further sub-tasks? If yes, in which ones?
- Are data collected at this point in the process? If yes, what data is collected, where is it sent (to which other process task), and where is it stored (at which storage location)?
- Are data used at this point in the process? If yes, what are these data used for, where were they previously collected (at which process task), and where are they stored (at which storage location)?
- Are the tools used correctly assigned? Are tools missing from the task?
- Are there any additional paths that can occur after this task that have not been depicted so far?

If questions cannot be answered or only partially answered, area experts can also be consulted here. Once the process has been walked through, the first interview ends. The PD processes all the information obtained during the interview and incorporates it into the model.

In the fourth process step, the process is validated through another interview with the process owner. In the second interview, the entire process is presented to the process owner again. If the process owner has no comments, the interview and the first validation unit are completed. However, if the process owner still has suggestions for improvements and corrections, the PD incorporates these into the model during the meeting. This avoids the need for further interview sessions with the process owner. The interview ends once the process owner perceives the model as complete and correct. Completeness and correctness also refer to the capturing of data and machine connectivity.

The second validation unit comprises a workshop with the relevant process participants. As outlined in Section 5.4.2, several relevant process participants are identified and invited to the workshop. In this workshop, the process model previously created with the process owner's assistance is reviewed. The workshop follows the structure of the last workshop presented in Section 5.4.2. This is to ensure that all parties involved agree on the completeness and correctness of the model, including data collection and connectivity. Upon completion of this workshop, there should be a process model that is highly likely to be correct and complete. Figure 5 shows the combined process discover process for IIoT-processes in BPMN.

# 7 Use Cases: Discover IIoT-Processes by Combining the Guidelines

This chapter presents the case studies conducted. The guidelines we proposed were applied in two independent case studies. Both case studies are described in detail below and are part of the evaluation in Section 8.

## 7.1 Use Case: Automotive Industry

In the first case study, we analyzed a manufacturing process from the automotive industry by ourselves combining parts of our guideline. The application took place as part of the evaluation in order to possibly initiate a further phase in the DSR cycle.

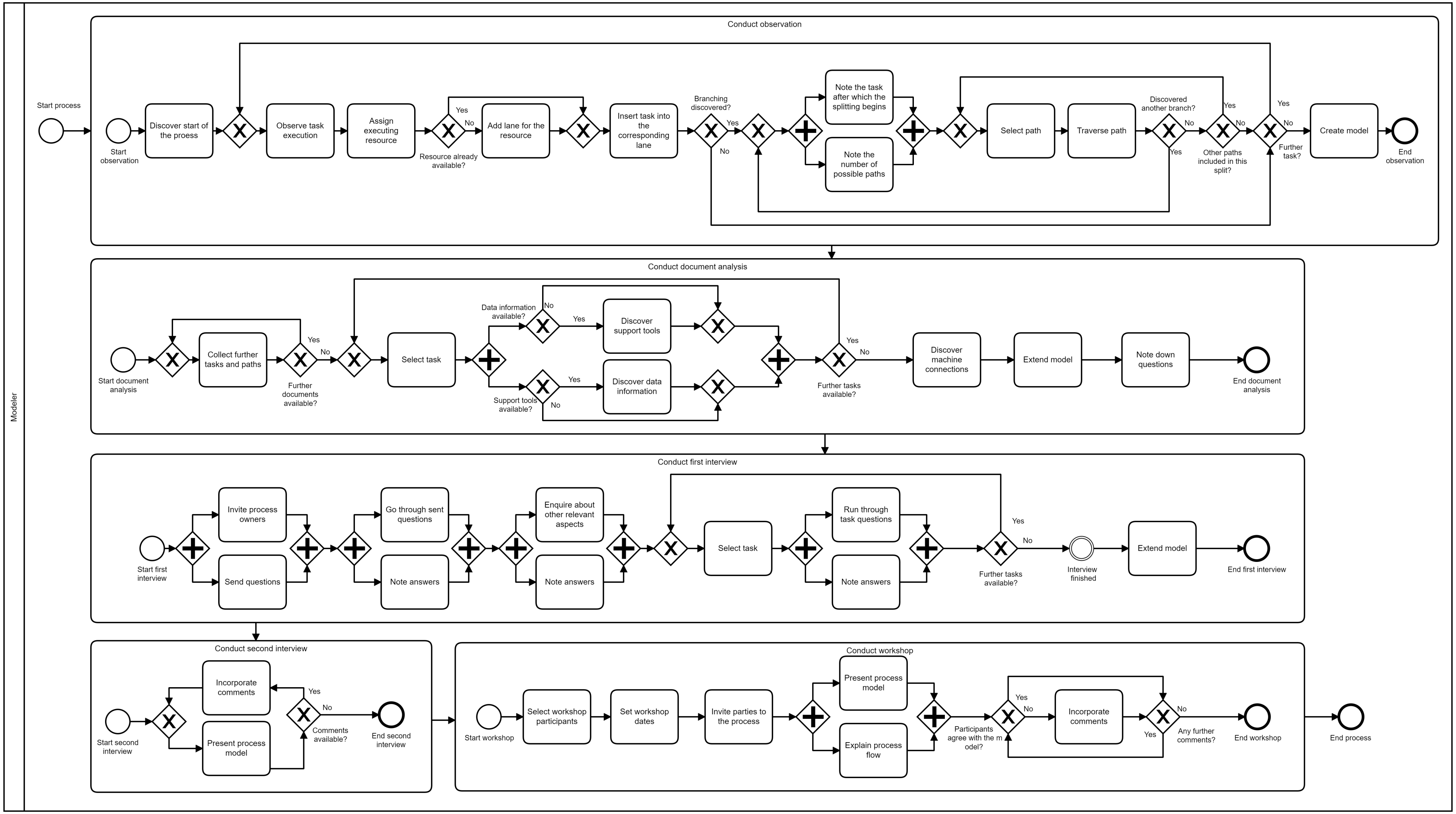

Figure 5: Process discovery for IIoT processes, presented in BPMN.

The name of the company remains anonymous for competitive reasons. The process model created can only be published in red. The production involves the manufacture of a component for automobiles and is therefore a production process where several machines and human actors are involved. The component produced is made up of 27 individual components, which are assembled in six higher-level, sequential process steps (bending, welding, assembly, moulding, testing, packaging).

The process discovery was started with a document analysis. This was followed by several interviews and subsequent observation combined with a workshop for validation purposes.

**Document Analysis - Preparation**

The process discovery began with document analyzes, as these were only available to a limited extent. The PD used the document analysis guideline to gain an initial insight into the process and defined questions for the interview based on the documents. The questions followed a top-down approach and were initially kept generic without going into process details in depth. The questions could therefore be taken from the suggestions for the questionnaire prior to conducting interviews.

**Interview - discover resources, tasks and work flow**

In a first, closed interview, the process owner enquired about the resources involved, their tasks and the process flow. The PD explained his understanding of the process and noted the corrections made by the process owner. After the interview, the PD then modelled an initial model based on the information provided. The PD noted open questions and ambiguities that arose during the modelling.

**Interview - discover data and tools**

The PD discussed the questions in a second interview with the process owner using the previous model. Once the questions had been clarified, the tools and data used in the process were discovered. It turned out that the connectivity and data connection could not be explained by any of the process participants, who were consulted as area experts. The process owner was then asked in an open interview whether there were any other process components that had not been addressed yet. The information from the second interview was then integrated into the model by the PD.

**Interview - Validation**

The third and final interview was used for process validation. The PD discussed the model with the process owner task by task and was assured that the sequence, tasks, tools, data and resources were correct and complete. The process owner's comments and corrections were incorporated directly into the model during the interview.

**Observation and Workshop - Validation**

For validation, the PD then observed the process on site. This was combined with a workshop in which the PD had employees explain to him exactly what took place in the individual process steps observed. He compared this with the model in parallel. Observation of the process on site showed that the results of the interviews already corresponded almost completely to the actual process flow. The chronological order was completely consistent. All process steps in the interview could be seen in the observation. There were no errors in content (e.g. incorrectly described tasks). The process steps that could not be recorded using the interview amount to a small number of five tasks in relation to the overall size of the model. Only three of these tasks were relevant to the process.

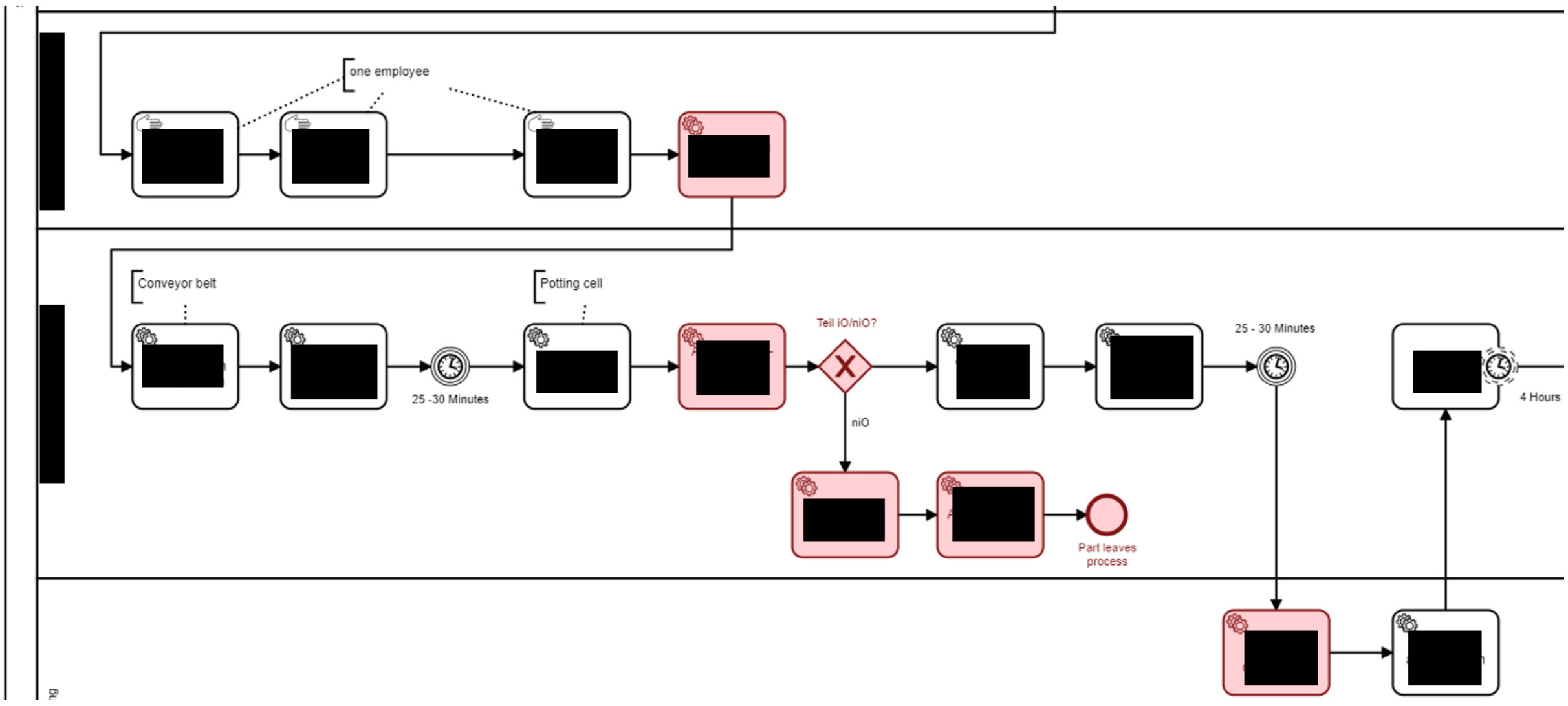

Figure 6: Red tasks were detected during validation.

The resulting model can be seen redacted in Figure 12 in the Appendix. The inserted tasks after validation are also shown in Figure 6. The model created does not contain any gaps or incomplete paths. Based on the validation, it can be assumed that the model is correct and complete. The granularity of the process could be deepened. Figure 13 in the appendix shows the process model of the autostore process after the observation.

## 7.2 Use Case: Warehouse Management

In a second use case, the "autostore" warehouse management process was discovered at SOMIC, a company specializing in the manufacturing of packaging machines. The process includes warehouse management using mobile robots that take required components from the stored boxes according to a production needs and transport them to the output portal. The robots travel on an aluminium construction that serves as rails to remove the individual bins from their container stacks as required.

The guidelines were handed out to three novices with no prior knowledge of process discovery. They discovered the process using the following procedure:

**Preparation**

The preparation began with contacting the company and making an appointment for the observation and an interview. In addition, process documents for the document analysis and employees for the workshop were requested. The questionnaire for the interview was already sent along. As no process documents were available and due to a lack of time, the document analysis and workshop were not carried out.

**Observation**

The observation was carried out independently of other discovery methods. For this purpose, each process step was observed individually in the forward observation and recorded in tabular form, as can be seen in Table Table 6. As a result, all perspectives were discovered during the observation of a task. The modelling was carried out simultaneously. Subsequently, a backward observation was carried out in which the observed tasks were compared with the model.

| Task | Number of triggers | Associated resource | Associated tools | Previous task | Subsequent task | Modelling status |
|---|---|---|---|---|---|---|
| **01 - Start Commissioning** | 1 | R04 | Order list | | 02 | x |
| **02 - Accept order** | 1 | R04 | Label | 01 | 03 | x |
| **03 - Robot starts** | 1 | R01, R02 | | 02 | 04a/04b | x |

Table 6: Table to handle process observation informations.

**Interview**

To conduct the interview, a questionnaire was sent to the process owner in advance, but this could not be completed due to a lack of time. The questionnaire was therefore discussed in the first interview, in which the resources, tasks, control flow, data and tools were enquired about. At the same time, an initial model was created. In the second interview, details were provided and queries clarified. The model created, which was collected independently of the previous observation, can be seen in the appendix in Figure 14. Finally, the model was sent to the process owner by email for validation.

**Combination**

Finally, the findings from the observation and the interview were combined. The result was that new tasks and a resource could be added to the observation. Overall, this resulted in a more detailed model, which can be seen in Figure 15 in the appendix.

The model created does not contain any gaps or incomplete paths. Due to the subsequent merging, it can be assumed that the model is correct and almost complete. The granularity of the process could be deepened. The IIoT perspective was clarified via the integrated database, which is linked to the respective tasks. The aim is to organise a workshop to possibly complete the model.

# 8 Evaluation

Three approaches were used for the evaluation. One is the observable approach proposed by Hevner et al. (2004): *(i)* in two case studies, the guidelines were used as artifacts in the business environment as described in Section 7, *(ii)* a survey on the necessity was conducted and *(iii)* focus groups discussed the presented guidelines.

We evaluate the guidelines from two perspectives: On the one hand, we want to analyze whether the guidelines are understandable and usable, based on the criteria of Prat et al. (2015). The artifact itself should therefore be examined. To this end, the artifact was used in two case studies that users can assess its understandability and usability. Furthermore, the guidelines were discussed in the focus groups with regard to their understandability and usability

On the other hand, the result of the guidelines should be checked. The intention is to ensure that models modelled using the guidelines are correct and complete. Models that were discovered using the guidelines cannot be compared with another, already existing model of the same process. This is because no other discovery techniques can be used except those that are to be evaluated. Process mining techniques cannot be used either due to the partial automation. Nevertheless, to be able to make an as-sessment, the information discovered in the case studies was translated into models and these were then analyzed by validation steps.

| Level of Experience / Environment | **Novices** | **Beginner** | **Advanced** | **Expert** |
|---|---|---|---|---|
| **Business and Economy** | | | Focus group 1 | |
| **Science** | Use Case 2 | Focus group 2 | | |

Table 7: Different user groups for evaluation approaches.

To look at the guidelines in detail and from different perspectives, we included different user groups with different levels of knowledge in the evaluation alongside three different techniques. Table 7 shows the different user groups and their distribution across the evaluation approaches.

Based on DSR, we evaluated the developed guidelines in several cycles using various approaches during their creation. These can be seen in Figure 7. The figure shows the results of the various evaluation approaches.

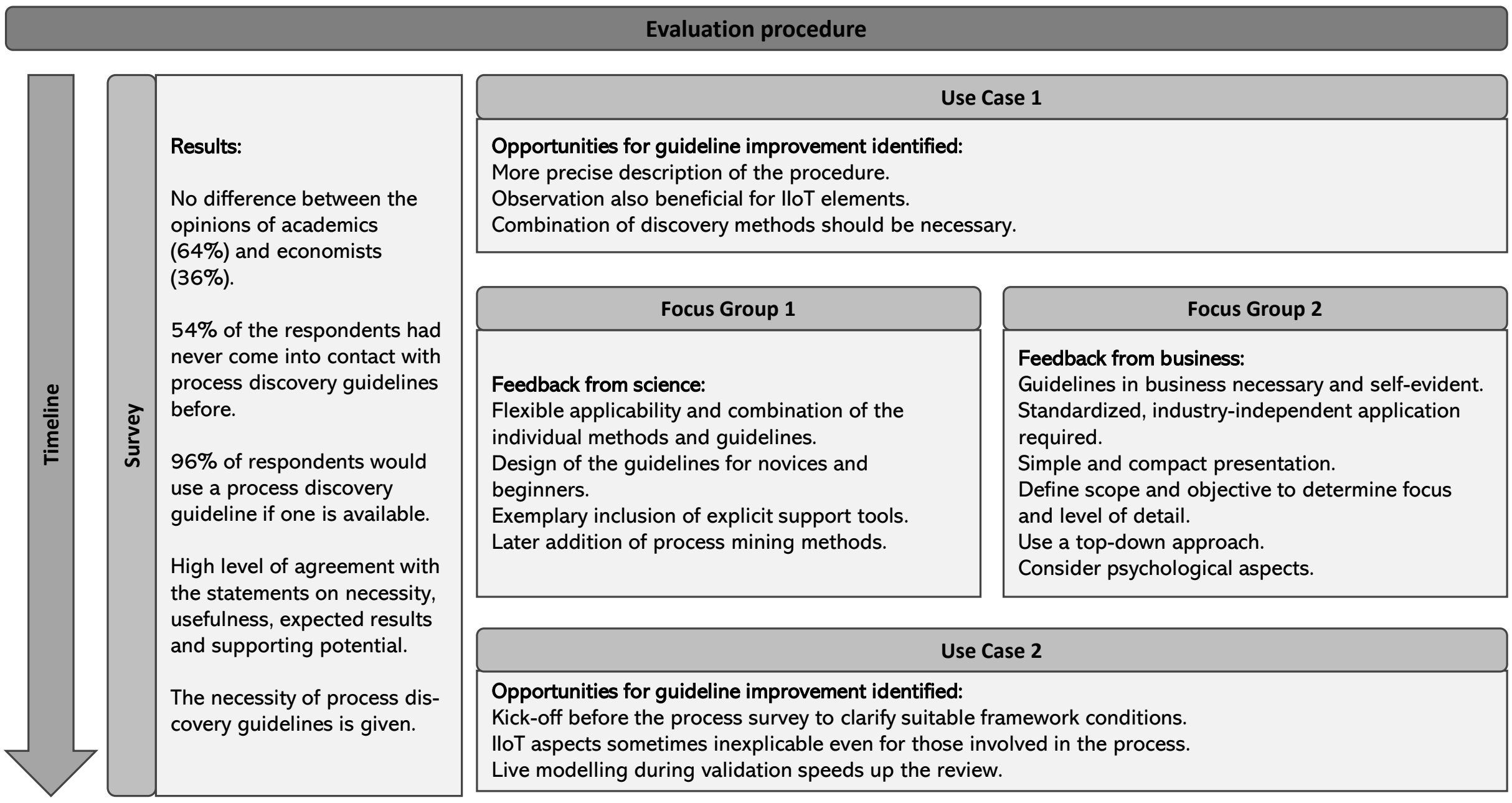

Figure 7: Results of the various evaluation approaches to improve the guidelines.

**Survey Results** The first step was to motivate the necessity and the advantages of guidelines in the process discovery. For this purpose, we developed a questionnaire based on the criteria of Prat et al. (2015). The questionnaire was designed based on Laugwitz et al. (2008) We conducted this survey with 28 participants from academia and industry. Participants of all levels of experience were invited to take part. Five questions at the beginning are intended to capture the background and environment of the interviewees with regard to their experience of process discovery. Subsequently, it is explicitly asked whether guidelines for process discovery would be used if they were available. The participants are asked to give their assessment of the given statements regarding the necessity, the usefulness, the expected result and supporting potential of process discovery guidelines. Each statement can be rated with *"Strongly agree"*, *"Agree"*, *"Neither agree nor disagree"*, *"Disagree"* or *"Strongly disagree"*. The exact questionnaire and its results can be seen in the Figures 8 to 11.

We have defined three hypotheses that we want to test using the survey:

**H1:** Guidelines would be used by people regardless of their professional background, experience and whether they have already conducted a process discovery.

**H2:** Guidelines are necessary to avoid errors and to ensure that all relevant information has been collected.

**H3:** Guidelines offer support potential to save time and costs and increase efficiency.

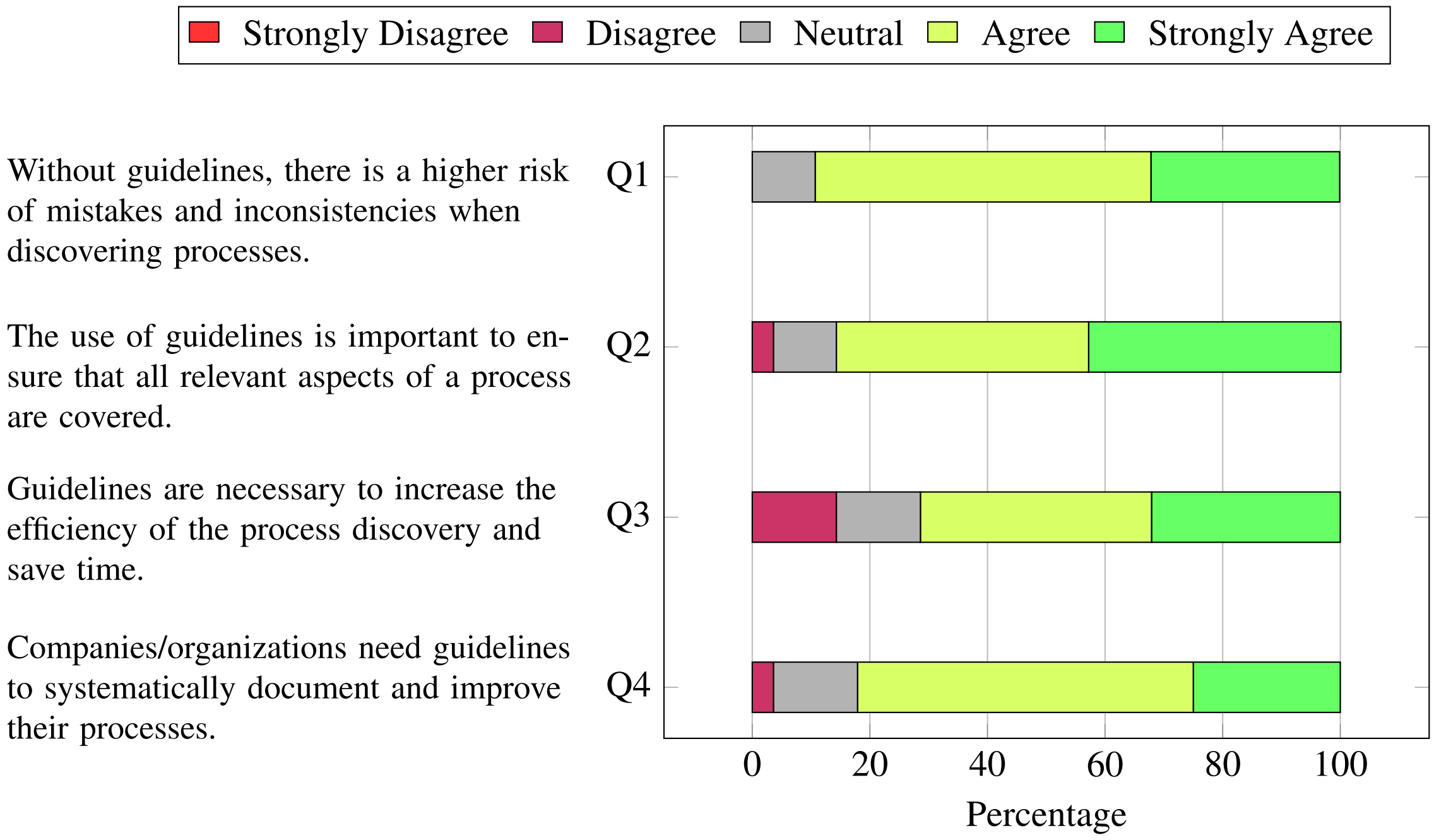

Figure 8: Questions and percentages of the necessity.

The following results of the study should be emphasized:

- Based on the results, we found that there was no difference between the opinions of academics (64%) and economists (36%).
- The majority of respondents classified themselves as "Advanced" (46%). 25% said they were "Experts", 21% "Beginners" and 7% with "No Experience".
- 82% of respondents had already completed a process discovery, including all experts, advanced and some beginners.
- Over half of respondents (54%) had never come into contact with process discovery guidelines before.
- The process discovery methods "document analysis", "interview" and "workshop" are equally well known (26% each). Only "Observation" was known by fewer respondents (16%).
- 96% of respondents stated that they would use a process discovery guideline if one was available.

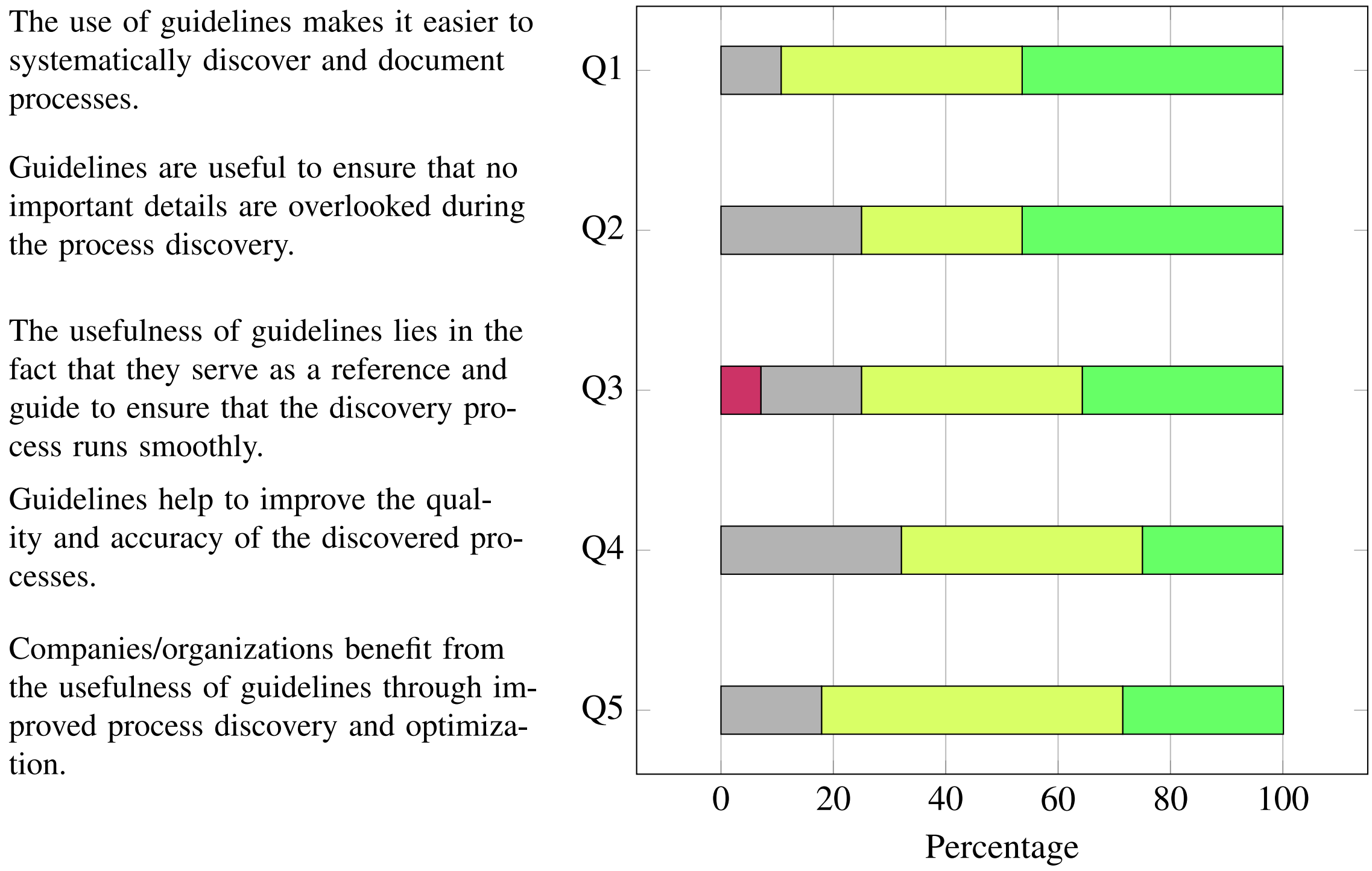

Figure 9: Questions and percentages of the usefulness.

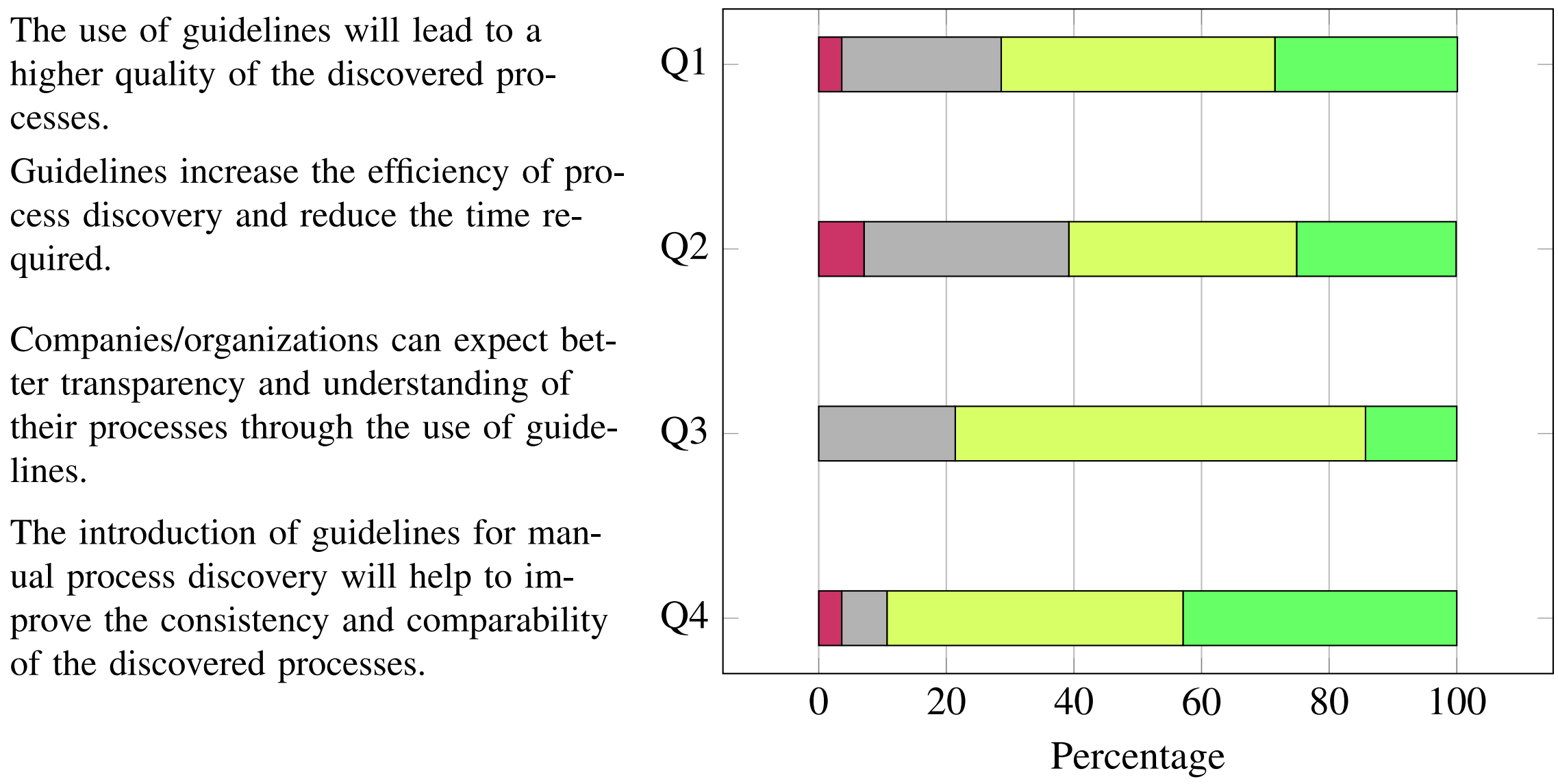

Figure 10: Questions and percentages of the expected result.

Viewing the results, we were able to summarise that the necessity an additional benefit of process discovery guidelines is given. In addition, guidelines improve the process survey in various aspects. The survey shows the independent consensus of expert opinions. This confirms **H1**. Due to the high level of agreement with the statements made, **H2** can be regarded as confirmed. It should be emphasised that none of the statements made were contradicted with a "strongly disagree". The most frequent "disagree" occurred with statements that support **H3**, time and cost reduction. Nevertheless, these statements were also regarded as given by a majority.

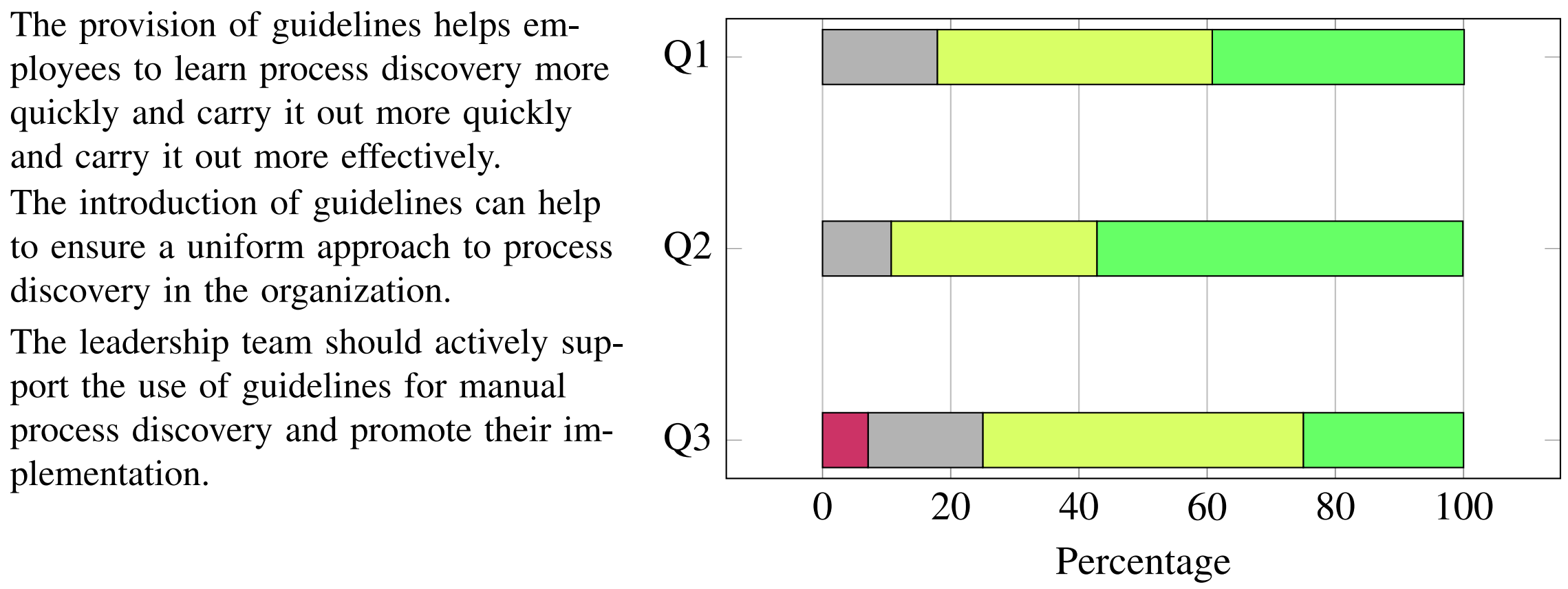

Figure 11: Questions and percentages of the supporting potential.

**First Use Case** Our second evaluation step was to use our approach in a first use case. This was carried out at an automotive supplier, as described in Section 7.1. After the implementation, it was determined that a more precise description of the procedure should be provided, as there were still unanswered questions regarding the specific discovery methods. Furthermore, it was found out that observation can also be beneficial for process discovery in IIoT processes, as it only became apparent during the discovery at which points in the process the component ID was scanned and the data sent to the control system. The scanners could be physically observed and therefore recorded directly.
Finally, it became clear that a combination of methods is necessary in order to be able to carry out process discoveries more flexibly on the basis of existing conditions. For example, it turned out that observation was particularly suitable for recording the control flow, while the description of the tools used was better captured by the interview.

**First Focus Group** After the first use case, the adapted guidelines were discussed in a focus group with people from the scientific community who have a theoretical background in process discovery and have already come into contact with process discovery. The discussion focused on the interpretation and target group of the guidelines. It was established that the guidelines must allow for flexible applicability in order to be able to balance out the advantages and disadvantages of the individual guidelines in each case with other approaches. The guidelines can be used by novices, advanced process explorers and experts alike. The guidelines are designed in particular for people with little or no experience in process discovery.
In addition, the exemplary inclusion of explicit support tools such as checklists, possible questions and further literature was requested. The person-centred discovery was critically examined, which is to be enriched in a next step with automated methods of process mining as a further possible method. Finally, concrete suggestions for improvement were made for the respective guidelines, which were then reviewed and integrated into the guidelines.

**Second Focus Group** In a further focus group, experts in the practical application of process dis-covery were invited. The experts confirmed the existence of individual, customized process discovery guidelines in corporations and their necessity. It was reported that small and medium-sized companies in particular regularly ask consulting firms for help with process discovery and methodological support, as there are no known procedures for standardized methods. Improvements and additions were made to these guidelines based on practical experience. Overall, it was pointed out that the guidelines should be clearly structured and compact. Uniform application should be possible regardless of the sector.
It was pointed out that the user should be made aware of the previous design of the process map in order to be able to classify processes correctly in context. In addition, the scope and objective of the process discovery should be defined before the discovery begins that it is clear which perspectives need to be focused on and how detailed the discovery should be. It is advisable to use a top-down approach and to detail the model step by step. Reference should be made to psychological aspects of the process discovery in interviews and workshops, but also during observation, as these are essential for discovery.

**Second Use Case** Our last evaluation step was to use our approach in a second use case. The application of the guidelines in the second use case is described in Section 7.2. The following results were obtained:
The choice of observation period can affect the success of the observation, so it is recommended to organise a kick-off before the observation to determine a suitable time window with the responsible parties. The detection of IIoT features is limited and there is overall uncertainty as to whether the model is complete, so a different method of validation should be chosen.
The interview revealed difficulties in answering the questionnaire, which is why several sector experts had to be consulted. One consideration is to supplement the interview with the workshop. In addition, some IIoT-relevant process components turned out to be a "black box" as they were operated by external service providers and there was therefore no insight or understanding of them. One possibility is to ask about the expertise of those involved in the process in the questionnaire at the beginning in order to reduce the time required and to be able to contact informed departmental experts directly. Finally, it was found that live modelling during validation provides more comprehensive feedback than a survey.
A combination of methods led to more findings, but also increased the discovery time. The sequence of the methods used should therefore be checked and determined sensibly before discovery.

# 9 Limitations and Future Research

This chapter discusses the limitations of the work and offers an outlook on further development ideas for future research in this area.
A major limitation of the guidelines is that no guarantee can be given that the discovery of a process is fully accurate and complete. This is due to the increasing complexity and size of processes, particularly in industry. In addition, it is not possible to compare the processes discovered with reference processes. Furthermore, the increasing complexity, especially due to automation and digitization, makes it increasingly difficult for employees to know and understand a process completely and in detail. It can be assumed that a large number of experts will be required to analyze the process, especially the larger the process and the more automated components it contains. In our use cases and focus groups, we found that the understanding of technical workflows and networks within the process was usually not documented and no contact person could be found who could answer questions in this regard. It therefore seems difficult to survey all IIoT-relevant process components.
In order to remedy these problems, the present work can be further developed and improved in several directions:

**Integration of Process Mining**

In order to process discovery, particularly on IIoT-relevant process components, the discovery methods should be expanded in future by integrating process mining techniques. In this way, the models for the automated process sections could be created using process mining, while the manual process sections would continue to be recorded using traditional or customised methods. A fusion of these two approaches would have the potential to simplify the time-consuming collection process for IIoT-specific processes while increasing the accuracy and completeness of the model.

**Automation of Document Analysis and Observation**

Another research approach is to extend the methods developed by using suitable tools. One possibility would be to automate document analysis by developing a tool that extracts initial process information from scanned or digital documents and creates a preliminary process model. In the case of documents that have already been digitised, this tool could speed up the creation of a preliminary process model. The PD has to go through the documents in the described iterations in order to *(i)* check the correctness of the process model and *(ii)* make additions that were not captured by the tool. The more documents that are already digitised and can therefore be pre-processed automatically, the more time could be saved. There is already research in this area, such as the automated "Rapid Process Discovery" procedure. This uses template-based filtering to extract patterns and information-based filtering to extract specific process information from the documents Ghose et al. (2007). Furthermore, Ackermann et al. (2021) and Neuberger et al. (2023) are currently developing approaches and tools that can generate models based on process descriptions. Their research includes natural language approaches.
Ideas for automating the observation process are proposed by Fichtner et al. (2020). If process models can be generated by analysing image and video material, it would be possible to automate the observation itself. Similar to document analysis, preliminary models could be created on the tool side using the material, which then have to be validated and supplemented by the process analyzer during the observation.

# Appendix A Extended Data

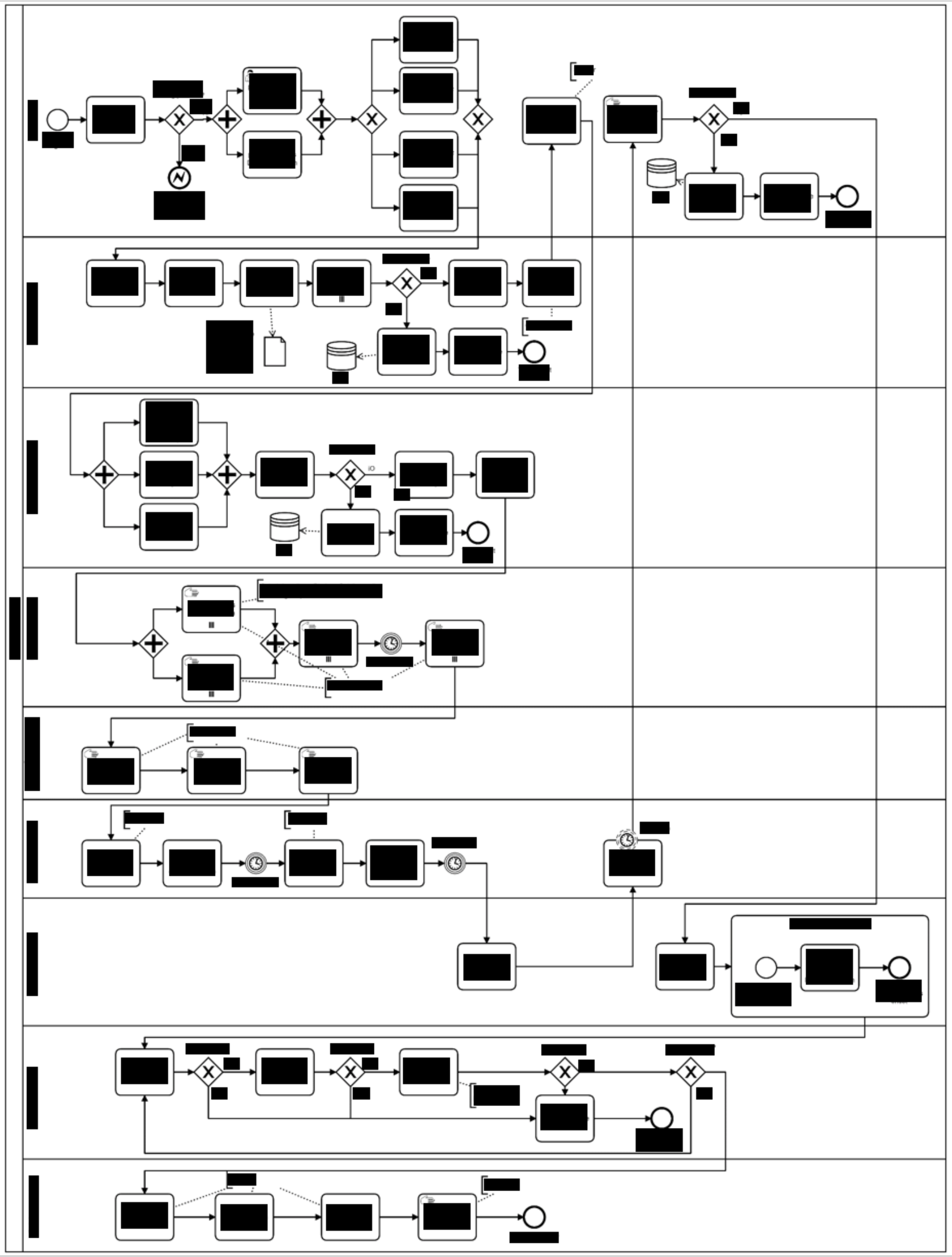

Figure 12: Blackened industrial process for the manufacture of engine components.

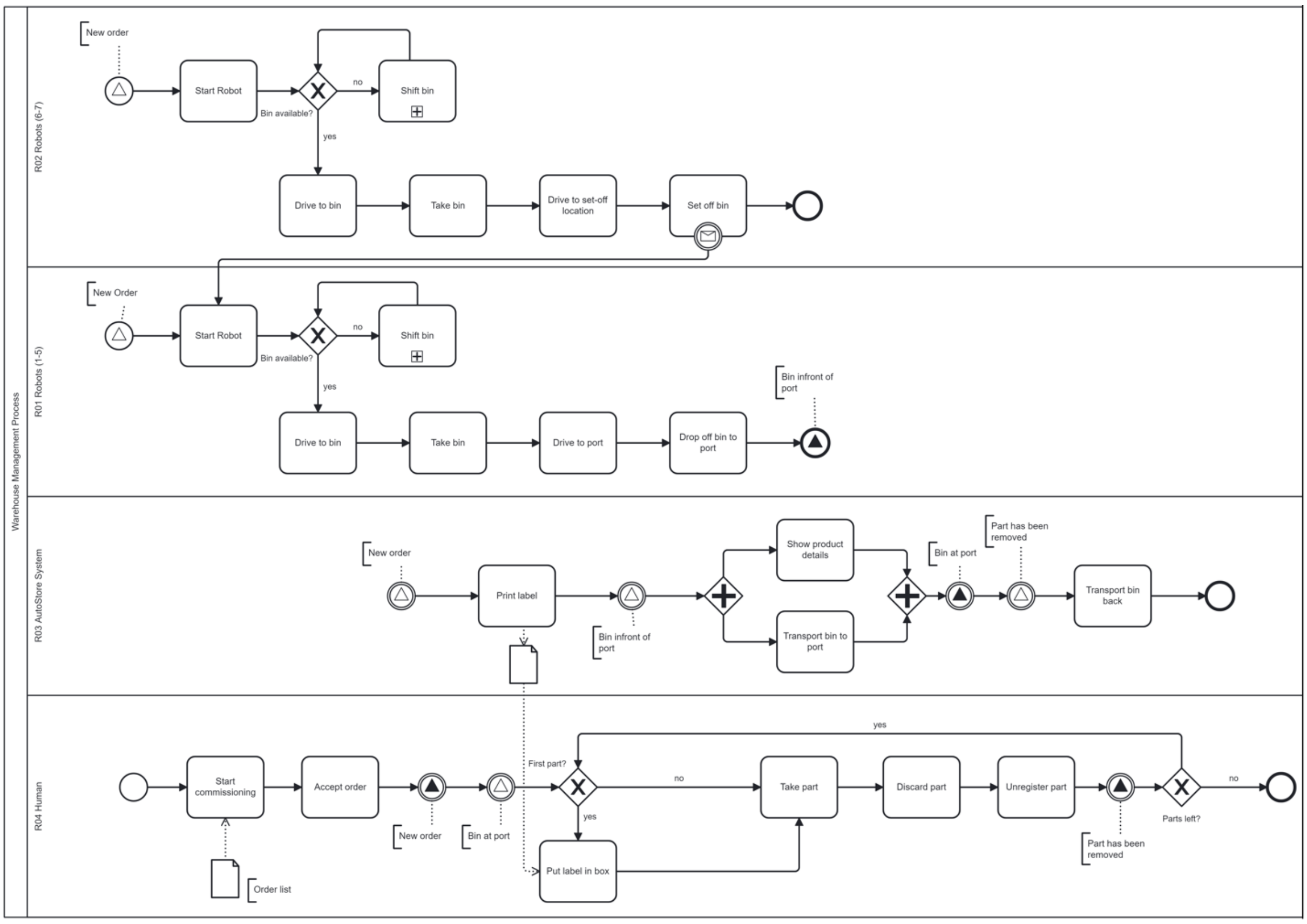

Figure 13: Model of the autostore process at Somic after observation.

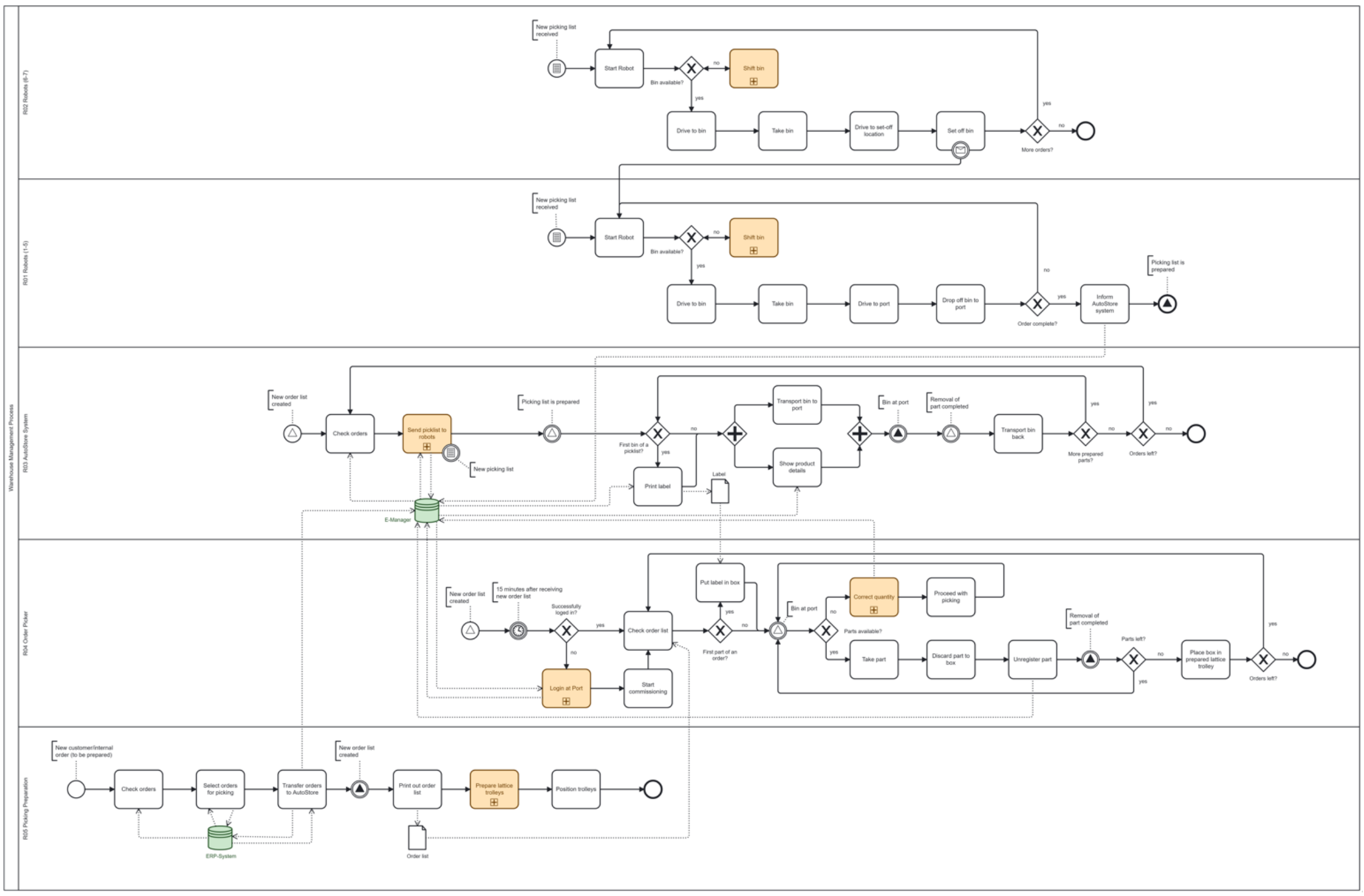

Figure 14: Model of the autostore process at Somic after interview.

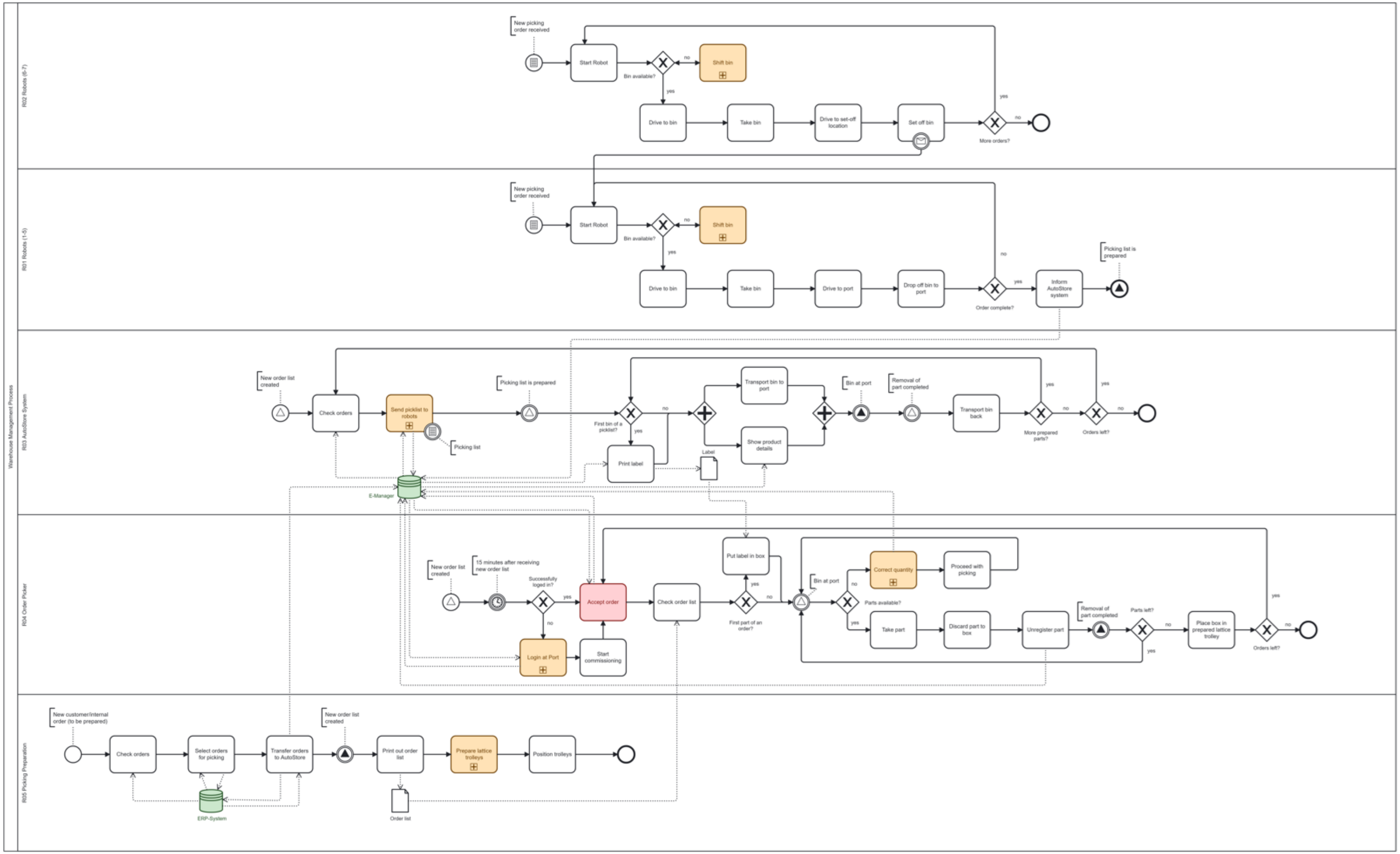

Figure 15: Model of the autostore process at Somic after combining observation and interview.